\documentclass[twocolumn,twoside]{IEEEtran} %!PN

\usepackage{amsmath,amssymb,amsfonts}
\usepackage[final]{graphicx}
\usepackage{psfrag}
\usepackage[numbers,sort&compress]{natbib}
\usepackage{flushend}

\newtheorem{theorem}{Theorem}
\newtheorem{proposition}{Proposition}

\newtheorem{definition}{Definition}

\def \A {{\mathcal A }}

\def \P {{\mathcal P }}
\def \I {{\mathcal I }}
\def \J {{\mathcal J }}

\def \H {{\mathcal H }}
\def \cov {\textit{cov}}
\def \R {\bar{\bf R }}

\def \M {{{\mathcal{MI}}}}

\makeatletter
\def\blfootnote{\xdef\@thefnmark{}\@footnotetext}
\makeatother

\begin{document}
\title{\Huge{Eigenvalue Dynamics of a Central Wishart Matrix with Application to MIMO Systems}}

\author{\IEEEauthorblockN{F. Javier Lopez-Martinez,~\IEEEmembership{Member,~IEEE},
Eduardo Martos-Naya,
Jose F. Paris,
Andrea Goldsmith,~\IEEEmembership{Fellow,~IEEE}}
}
%\markboth{IEEE Transactions on Information Theory}
%{Lopez.Martinez \MakeLowercase{\textit{et. al}}: Eigenvalue Dynamics of a Central Wishart Matrix}

\maketitle

\begin{abstract}
We investigate the dynamic behavior of the stationary random process defined by a central complex Wishart (CW) matrix ${\bf{W}}(t)$ as it varies along a certain dimension $t$. We characterize the second-order joint cdf of the largest eigenvalue, and the second-order joint cdf of the smallest eigenvalue of this matrix. We show that both cdfs can be expressed in exact closed-form in terms of a finite number of well-known special functions in the context of communication theory. As a direct application, we investigate the dynamic behavior of the parallel channels associated with multiple-input multiple-output (MIMO) systems in the presence of Rayleigh fading. Studying the complex random matrix that defines the MIMO channel, we characterize the second-order joint cdf of the signal-to-noise ratio (SNR) for the best and worst channels. We use these results to study the rate of change of MIMO parallel channels, using different performance metrics. For a given value of the MIMO channel correlation coefficient, we observe how the SNR associated with the best parallel channel changes slower than the SNR of the worst channel. This different dynamic behavior is much more appreciable when the number of transmit ($N_T$) and receive ($N_R$) antennas is similar. However, as $N_T$ is increased while keeping $N_R$ fixed, we see how the best and worst channels tend to have a similar rate of change.
\end{abstract}

%\clearpage
\blfootnote{Manuscript received June 19, 2014; revised October 11, 2014; accepted
February 24, 2015. This work was supported in part by Junta de Andalucia (P11-TIC-7109), Spanish Government-FEDER (TEC2010-18451, TEC2011-25473, TEC2013-44442-P, COFUND2013-40259), the University of Malaga and European Union under Marie-Curie COFUND U-mobility program (ref. 246550),  NEC, Huawei, and the NSF Center for Science of Information.\\This manuscript was presented in part at IEEE Information Theory Workshop 2013 \cite{Lopez2013b}.\\ \indent F. J. Lopez-Martinez is with Dpto. Ingenieria de Comunicaciones, Universidad de Malaga, Spain. He previously was with the Wireless Systems Lab, Department of Electrical Engineering, Stanford University, CA, USA. (email: fjlopezm@ic.uma.es)\\ \indent A. Goldsmith is with the Wireless Systems Lab, Department of Electrical Engineering, Stanford University, CA, USA. (email: andrea@wsl.stanford.edu).\\ \indent E. Martos-Naya and J. F. Paris are with Dpto. Ingenieria de Comunicaciones, Universidad de Malaga, Spain. (email: eduardo@ic.uma.es, paris@ic.uma.es)\\
\indent Copyright (c) 2014 IEEE. Personal use of this material is permitted.  However, permission to use this material for any other purposes must be obtained from the IEEE by sending a request to pubs-permissions@ieee.org.}
\section{Introduction}
%\blfootnote{This work has been submitted to the IEEE for possible publication. Copyright may be transferred without notice, after which this version may no longer be accessible.}
\subsection{Related work}
Since the seminal work by Wishart \cite{Wishart1928}, random matrix theory has found application in very diverse fields like physics \cite{Guhr1998}, neuroscience \cite{Rajan2006} and many others \cite{Akemann2011}. For instance, random matrix processes are useful in econometrics to study the stock volatility in portfolio management \cite{Gou2009,Rubio2012}; in immunology, random matrix theory has been used to design immunogens targeted for rapidly mutating viruses \cite{imm2011}.

In the context of information and communication theory, random matrices have been used to characterize the performance of communication systems that make use of multiple antennas at the transmitter and the receiver sides of a communication link, referred to as a multiple-input multiple-output (MIMO) systems. This technique has become the standard transmission mechanism for current wireless communication systems \cite{Goldsmith2005} due to its increased capacity and reliability. In this scenario, the channel is described as a random matrix ${\bf{H}}$ whose size is determined by the numbers of transmit and receive antennas.

The characterization of the eigenvalues of the matrix ${{\bf{W}}\triangleq{\bf{HH^{\H}}}}$ has been used to study the fundamental performance limits of MIMO systems \cite{Foschini1996,Telatar1999}; specifically, the ordered eigenvalues of ${\bf{W}}$ characterize the parallel eigenchannels used to achieve multiplexing gain, and, in particular the largest eigenvalue of ${\bf{W}}$ determines the diversity gain of the system.

When the entries of ${\bf{H}}$ are distributed as complex Gaussian random variables, then ${\bf{W}}$ is said to follow a complex Wishart (CW) distribution \cite{Wishart1928}. The eigenvalue statistics of CW matrices have been studied in depth in the literature, both for central \cite{Khatri1964,Edelman1989,Ratnarajah2005,Chiani2014} and non-central \cite{Kang2003,Maaref2007,Jin2008,Ordonez2009,Zanella2009,Dharmawansa2011,Ghaderipoor2012} Wishart distributions. These results can be seen as a first-order characterization of a CW random process, and can be used to derive useful performance metrics such as the outage probability or the channel capacity.

However, wireless communication systems are in general non-static and hence the stochastic process associated with ${\bf{H}}$ exhibits a variation along different dimensions due to mobility of users or objects in the propagation environment. For example, temporal variation of wireless communication channels due to user mobility has an impact on the ability to estimate channel state information and hence limits the achievable performance. Equivalently, the frequency variation due to the effect of multipath has a similar effect on the equivalent channel gain in the frequency domain.

If we consider two samples of a stationary CW random process ${\bf{W}}(t)$, namely ${{\bf{W}}_1\triangleq{\bf{W}}(t)}$ and ${{\bf{W}}_2\triangleq{\bf{W}}(t+\tau)}$, the dynamics of ${\bf{W}}(t)$ are captured by the joint distribution of ${{\bf{W}}_1}$ and ${{\bf{W}}_2}$. More precisely, the dynamics of the MIMO parallel channels (or \textit{eigenchannels}) can be studied separately by studying the joint distribution of the eigenvalues of ${{\bf{W}}_1}$ and ${{\bf{W}}_2}$. Along these lines, the statistical analysis of \textit{two correlated} CW matrices was tackled in \cite{Smith2007,Kuo2007}, deriving the $2s$-dimensional joint pdf of the $s$ eigenvalues of a CW matrix and a perturbed version of it. Rather than the \textit{joint distribution of all the eigenvalues}, we consider the \textit{marginal joint distribution} of a particular eigenvalue as our metric to capture the dynamic behavior of a CW random process, since this distribution allows for the separate statistical characterization of all $s$ eigenvalues. Therefore, we will focus our attention on this set of second-order (or bivariate) distributions.

This problem was addressed in \cite{McKay2008} when studying the mutual information distribution in orthogonal frequency division multiplexing (OFDM) systems operating under frequency-selective MIMO channels; specifically, a closed-form expression for the joint second-order pdf was given for arbitrarily-selected eigenvalues of the equivalent frequency-domain Wishart matrix. However, in order to obtain the joint bivariate cdf or the correlation coefficient for a particular eigenvalue, a two-fold numerical integration with infinite limits was required. In \cite{Kongara2009}, an expression for this bivariate cdf was derived for the extreme eigenvalues (i.e. the largest and the smallest) in terms of the determinant of a matrix whose entries are expressed as infinite series of products of incomplete gamma functions; hence, its evaluation is highly impractical as the number of antennas is increased.

\subsection{Contributions}
In this paper, we show that the joint second-order cdf of the random process given by the largest eigenvalue of a complex Wishart matrix can be expressed in closed-form; this also holds when considering the smallest eigenvalue. Specifically, we provide exact closed-form analytical results for these distributions in terms of the determinant of a matrix, whose entries are expressed as a finite number of Marcum $Q$-functions \cite{Marcum1950} and modified Bessel functions of the first kind. Therefore, they can be efficiently computed with commercial software mathematical packages.

These results complete the current landscape of closed-form bivariate cdfs for the most common fading distributions such as Rayleigh \cite{Schwartz1966}, Weibull \cite{Sagias2005} and Nakagami-$m$ \cite{Lopez2013,Morales2013}. Interestingly, they are expressed in terms of the Marcum $Q$-function, and hence the results for Nakagami-$m$ and Rayleigh fading can be seen as particular cases of the expressions derived herein when particularized for $N_T=m,N_R=1$ and $N_T=1,N_R=1$ respectively. 

We also show how our results can be used to directly evaluate many performance metrics that allow us to quantify the rates of change of MIMO parallel channels in different ways:
\begin{enumerate}
\item We study the behavior of a mutual information metric associated with two different observations of the eigenvalue of interest. We quantify the loss in mutual information as the CW process changes, in order to determine the impact of increasing $N_T$ or $N_R$ in the rate of change of the best and worst parallel channels.
\item We evaluate the probability of having two outages in two different instants. This metric applies to two different transmissions within a given time window $\tau$ in flat fading channels, as well as to two different transmissions with a frequency separation $\Delta f$ in MIMO OFDM systems affected by multipath fading. 
\item The level crossing rate (LCR) and the average fade duration (AFD) are often used to characterize the dynamics of fading channels. In the context of time-varying MIMO channels, this problem was tackled in \cite{Ivanis2007} using Rice's original framework for LCR analysis \cite{Rice1944} of continuous random processes. However, the inherent sampling of the equivalent frequency-domain channel in OFDM systems makes the LCR obtained using Rice's approach to be an overestimation of the actual LCR as observed in \cite{Kongara2011}. Here, we use our closed-form results for the bivariate cdfs of interest to study the LCR and the AFD of the extreme eigenvalues in MIMO OFDM systems, using the method described in \cite{Lopez2012} for sampled fading channels.
\end{enumerate}

The remainder of the paper is structured as follows: The main mathematical contributions are presented in Section \ref{Analysis}, i.e. the bivariate cdfs of the extreme eigenvalues of two correlated CW matrices. As an application, the performance metrics that characterize the dynamics of MIMO parallel channels are introduced in Section \ref{Metrics}, and then used in Section \ref{Applications} to provide some numerical results in practical scenarios of interest. Finally, our main conclusions are discussed in Section \ref{Conclusion}. The proofs for the results in Section \ref{Analysis} are included as appendices.

\section{Statistical analysis}
\label{Analysis}
\subsection{Notation and preliminaries}
Throughout this paper, vectors and matrices are denoted in bold lowercase $\bf{a}$ and bold uppercase $\bf{A}$, respectively. We use $|a|$ to indicate the modulus of a complex number $a$, whereas $|{\bf{A}}|$ indicates the determinant of a matrix $\bf{A}$. The symbol $\sim$ means \textit{statistically distributed as}, whereas $\text{E}\{\cdot\}$ represents the expectation operation, and the super-index ${}^{\mathcal{H}}$ denotes the Hermitian transpose. For the sake of coherence with the scenario where the general results of this paper are used, we denote the numbers of rows and columns of the random matrices with Gaussian entries as $N_R$ and $N_T$, respectively.

The correlation coefficient of two random variables $X$ and $Y$ is defined as
\begin{equation}
\label{powcorr}
{\rho_{X,Y}\triangleq \frac{\cov\{X,Y\}}{\sqrt{\sigma_{X}^2{\sigma_{Y}^2}}}},
\end{equation}
where $\cov\{\}$ denotes covariance operation, and $\sigma_{X}^2$,$\sigma_{Y}^2$ represent the variances of the random variables $X$ and $Y$ respectively.

The cdf of $X$ is defined as $F_X(x)\triangleq \Pr\{X\leq x\}=\int_{-\infty}^{x}f_X(z)dz$, where $f_X(x)$ is the pdf of $X$.
Similarly, the joint cdf of two correlated random variables $X$ and $Y$ is defined as
\begin{align}
\label{relcdfccdf}
{F_{X,Y}(x,y)}&\triangleq \Pr\{X\leq x,Y\leq y\}\\ \nonumber&=\int_{ - \infty }^x {\int_{ - \infty }^y {f_{X,Y} \left( {z_1 ,z_2 } \right)dz_1 dz_2 } },
\end{align}
where $f_{X,Y} \left( {x ,y }\right)$ is the joint cdf of $X$ and $Y$.
The joint complementary cdf (ccdf), defined as ${\bar F_{X,Y}(x,y)}\triangleq \Pr\{X> x,Y> y\}$ and the joint cdf of two correlated identically distributed random variables $X$ and $Y$ are related through
\begin{align}
\label{cdfccdf}
{\bar F_{X,Y}(x,y)}=F_{X,Y}(x,y)-F_X(x)-F_Y(y)+1.
\end{align}

\subsection{Definitions}
\label{Defs}
Before presenting the main analytical results, it is necessary to introduce the following definition of interest related with central CW matrices.
\begin{definition}
\label{def0}
\textit{Central Complex Wishart Matrix.}
\\
\textit{Let us consider the complex random matrix ${\bf{H}}\in \mathbb{C}^{N_R \times N_T}$ with zero-mean i.i.d. Gaussian entries $\sim\mathcal{CN}\left(0,\sigma^2\right)$. If we define $s=\min\left(N_R,N_T\right)$, $t=\max\left(N_R,N_T\right)$ and the matrix ${\bf{W}}$ as}
\begin{equation}
{\bf{W}} = \left\{ {\begin{array}{*{20}c}
   {{\bf{HH}}^H \,\,\,\,\,\,N_R  \le N_T }  \\
   {{\bf{H}}^H {\bf{H}}\,\,\,\,\,\,N_R  > N_T }  \\
\end{array}} \right.
\end{equation}
\textit{then ${\bf{W}}\in \mathbb{C}^{s \times s}$ follows a complex central Wishart distribution, i.e. ${\bf{W}} \sim \mathcal{CW}\left(t,\sigma^2{\bf{I}}_s,{\bf{0}}_s\right)$, where ${\bf{I}}_s$ and ${\bf{0}}_s$ are the identity and the null $s\times s$ matrices, respectively.}
\end{definition}

It is also necessary to introduce an auxiliary integral function.

\begin{definition}
\label{def2}
\textit{Incomplete integral of Nuttall $Q$-function}
\begin{equation}
\label{eqdef2}
\J^{\alpha ,\beta ,\gamma}_{k,c,j} \left(u\right) = \int\limits_u^\infty  {z^{2j + k - 1} e^{ - \alpha z^2 } Q_{2c + k + 1,k} \left( {\gamma z,\beta } \right)\:dz},
\end{equation}
\textit{where $Q_{m,n}(x,y)$ is the Nuttall $Q$-function \cite{Nuttall1974}, ${k,c,j \in \mathbb{Z}^+}$, ${u \in [0,\infty)}$ and ${\alpha,\beta,\gamma \in \mathbb{R}}$.}
\end{definition}

The function $\J^{\alpha ,\beta ,\gamma}_{k,c,j} \left(u\right)$ is an incomplete integral of Nuttall $Q$-function (IINQ), and generalizes a class of incomplete integrals of Marcum $Q$-functions which appear in different problems in communication theory \cite{Nuttall1974,Simon2005,Lopez2013}. 

We also restate previous results for the first-order statistics of the extreme eigenvalues of CW matrices that will be used in our later derivations.

\begin{proposition}
\label{th00a}
\textit{Let $\lambda_1$ be the largest eigenvalue of a complex central Wishart matrix ${\bf{W}} \sim \mathcal{CW}\left(t,\sigma^2{\bf{I}}_s,{\bf{0}}_s\right)$. The cdf of $\lambda_1$ is given in closed-form as \cite{Khatri1964,Kang2003}}
\begin{equation}
F_{\lambda_1}\left(\lambda\right)= \frac{\left|{\bf{M}}(\lambda)\right|}{\left|{\bf{K}}\right|} ,
\end{equation}
\textit{where}
\begin{align}
M_{i,j} &=\gamma(t-s+i+j-1,\frac{\lambda}{\sigma^2}),\\
K_{i,j} &=\Gamma(t-s+i+j-1),
\end{align}
$\gamma(m,x)$ \textit{is the lower incomplete Gamma function, and $\Gamma(\cdot)$ denotes the Gamma function}.
\end{proposition}

For the sake of compactness in the following derivations, we will use the ccdf of the smallest eigenvalue.

\begin{proposition}
\label{th00b}
\textit{Let $\lambda_s$ be the smallest eigenvalue of a complex central Wishart matrix ${\bf{W}} \sim \mathcal{CW}\left(t,\sigma^2{\bf{I}}_s,{\bf{0}}_s\right)$. The ccdf of $\lambda_s$ is given in closed-form as \cite{Khatri1964,Kang2003}}
\begin{equation}
\bar F_{\lambda_S}\left(\lambda\right)= \frac{\left|{\bf{\tilde M}}(\lambda)\right|}{\left|{\bf{K}}\right|} ,
\end{equation}
\textit{where}
\begin{align}
\tilde M_{i,j} =\Gamma(t-s+i+j-1,\frac{\lambda}{\sigma^2}),
\end{align}
\textit{and $\Gamma(m,x)$ is the upper incomplete Gamma function.}
\end{proposition}

\subsection{Problem Statement}
\label{sec:PS}
We are interested in the study of a stationary CW random process ${\bf{W}}(t)$. Hence, we consider two realizations of this random process at two different instants, i.e. ${\bf{W}}(t)\triangleq {\bf{W}}_1$ and ${{\bf{W}}(t+\tau)\triangleq {\bf{W}}_2}$. The correlation between the underlying Gaussian processes ${\bf{H}}(t)\triangleq {\bf{H}}_1$ and ${\bf{H}}(t+\tau)\triangleq {\bf{H}}_2$ corresponding to the two realizations of the Gaussian matrix can be modelled as
\begin{equation}
\label{eqcorr00}
{\bf{H}}_2 = \rho {\bf{H}}_1 + \sqrt {1 - \rho ^2 } {\bf{\Xi}},
\end{equation}
where $\rho$ is the correlation coefficient\footnote{In our case, we assume that each of the elements of this channel matrix ${h_{i,j}}(t)$ evolves as an independent random process. This is the natural extension of the i.i.d. case to include the evolution of the random matrix ${\bf{H}}$ as it varies along a certain dimension $t$. We also assume that the correlation coefficient for each of the channel matrix elements is the same, i.e. ${\bf{H}}_2 = {\bf{R}} {\bf{H}}_1 + {\R} {\bf{\Xi}},$  where ${\bf{R}}$ and $\R$ are $N_R\times N_R$ diagonal matrices with elements $r_{i,i}=\rho$ and $\bar{r}_{i,i}=\sqrt{1-|\rho|^2}$; therefore, both $\bf{R}$ and $\R$ have the form of the scaled identity matrix. As we will later see, this model is useful in a number of scenarios involving MIMO communications. Assuming a more general correlation model is indeed possible, although a different approach would be required in order to analyze the dynamics of the extreme eigenvalues, and a closed-form characterization is probably unattainable in such situation.} between the $\{i,j\}$ entries of ${\bf{{H}}}_1$ and ${\bf{{H}}}_2$, and ${\bf{\Xi}}$ is an auxiliary $N_R\times N_T$ matrix with i.i.d. entries $\sim\mathcal{CN}\left(0,\sigma^2\right)$, which is independent of ${\bf{{H}}}_1$.

In virtue of the spectral theorem, the matrices ${\bf{W}}_1$ and ${\bf{W}}_2$ are orthogonally diagonalizable, i.e. there exist $s\times s$ matrices ${\bf{V}}_1$ and ${\bf{V}}_2$ such as ${\bf{W}}_1={\bf{V}}_1{\bf{\Lambda}}{\bf{V}}_1^{\H}$ and ${\bf{W}}_2={\bf{V}}_2{\bf{\Phi}}{\bf{V}}_2^{\H}$. The diagonal matrices formed by the ordered eigenvalues of ${\bf{W}}_1$ and ${\bf{W}}_2$ are then given by ${{\bf{\Lambda}}\triangleq \text{diag}\{\lambda_1,...,\lambda_s\}}$ and ${{\bf{\Phi}}\triangleq \text{diag}\{\varphi_1,...,\varphi_s\}}$, where $\lambda_k$ and $\varphi_k$ represent the $k^{th}$ eigenvalue of the ${\bf{W}}_1$ and ${\bf{W}}_2$ matrices, respectively.

Throughout this paper, we will focus our attention on the extreme eigenvalues, i.e. the largest ($k=1$) and the smallest ($k=s$). Specifically, we are interested in the characterization of the dynamics of the random process given by the $1^{st}$ and $s^{th}$ eigenvalues of a complex Wishart matrix. Therefore, we aim to study the joint distributions of the random processes $\left\{\lambda_i,\varphi_i\right\}$ for $i=\{1,s\}$.

With the previous definitions, it is clear that the random matrices ${\bf{H_2}}|{\bf{H_1}}$ and ${\bf{W_2}}|{\bf{W_1}}$ are distributed as
\begin{align}
{\bf{H_2}}|{\bf{H_1}}&{\sim \mathcal{CN}_{N_R,N_T} (\rho {\bf{H}}_1,\sigma ^2 (1 - \rho ^2 ){\bf{I}}_{N_R}  \otimes {\bf{I}}_{N_T})}\\
\label{CWdist01}
{\bf{W_2}}|{\bf{W_1}}&{\sim \mathcal{CW}_s \left( {t,\sigma ^2 (1 - \rho ^2 ){\bf{I}}_s ,\tfrac{\rho ^2}{{\sigma ^2 (1 - \rho ^2 )}} {\bf{W}}_1} \right)}.
\end{align}
Hence, ${\bf{W_2}}|{\bf{W_1}}$ follows a non-central Wishart distribution with non-centrality parameter matrix given by ${\tfrac{\rho ^2}{{\sigma ^2 (1 - \rho ^2 )}} {\bf{W}}_1}$. The distribution of the $k^{th}$ eigenvalue of a non-central CW matrix was calculated in \cite{Jin2008}; here, we use the distribution of the eigenvalues for the conditioned random matrix ${\bf{W_2}}|{\bf{W_1}}$ to obtain exact closed-form expressions for the marginal joint distributions of the \textit{largest} and the \textit{smallest} eigenvalues of two correlated CW matrices.

\subsection{Main Results}
In the following theorem, we obtain the joint distribution of the largest eigenvalue of two correlated Wishart matrices.

\begin{theorem}
\label{th01}
\textit{Let $\lambda_1$ and $\varphi_1$ be the largest eigenvalues of two complex central Wishart matrices ${\bf{W}}_1$ and ${\bf{W}}_2$, respectively, where ${\bf{W}}_1$ and ${\bf{W}}_2$ are identically distributed ${\bf{W}} \sim \mathcal{CW}\left(t,\sigma^2{\bf{I}}_s,{\bf{0}}_s\right)$, with underlying Gaussian matrices correlated according to (\ref{eqcorr00}). The exact joint cdf of $\lambda_1$ and $\varphi_1$ can be expressed as}
\begin{equation}
\label{bivL}
F_{\lambda_1,\varphi_1}\left(\lambda,\varphi\right)=C  \left|{\bf{\Upsilon}}(\lambda,\varphi)\right|,
\end{equation}
\textit{where}
\begin{align}
C &=2^s \sigma ^{ - 2ts} ( - 1)^{\frac{{s(s - 1)}}{2}} \varepsilon^{\frac{{s(s - t)}}{2}} \prod\limits_{i = 1}^s {\tfrac{{\varepsilon^{1- i} }}{{\left( {s - i} \right)!\left( {t - i} \right)!}}},
\end{align}
\textit{$\kappa={\tfrac{{1}}{{(1 - \rho ^2 )\sigma ^2 }}}$, $\varepsilon=\kappa\rho^2$, the entries of the $s\times s$ matrix $\bf{\Upsilon}$ are given by}
\begin{align}
\Upsilon_{i,j}=&2^{(2i-s-t)/2}\left\{\J^{\sigma^{-2} ,0 ,\sqrt{2\varepsilon}}_{t-s,s-i,j} \left(0\right)-\J^{\sigma^{-2} ,\sqrt{2\kappa\varphi} ,\sqrt{2\varepsilon}}_{t-s,s-i,j} \left(0\right)\nonumber\right.\\ &\left.-\J^{\sigma^{-2} ,0 ,\sqrt{2\varepsilon}}_{t-s,s-i,j} \left(\sqrt{\lambda}\right)+
\J^{\sigma^{-2} ,\sqrt{2\kappa\varphi} ,\sqrt{2\varepsilon}}_{t-s,s-i,j} \left(\sqrt{\lambda}\right)\right\}.
\end{align}
%\textit{and $\upsilon_{i,s,t}=2^{(2i-s-t)/2}$}. 
\end{theorem}

\begin{IEEEproof}
See Appendix \ref{App1}.
\end{IEEEproof}

Following a similar approach, we derive a closed-form expression for the joint distribution of the smallest eigenvalue in the following theorem. In this case, and for the sake of compactness, we present an expression for the bivariate ccdf, being the cdf obtained in a straightforward manner using (\ref{cdfccdf}).

\begin{theorem}
\label{th01b}
\textit{Let $\lambda_s$ and $\varphi_s$ be the smallest eigenvalues of two complex central Wishart matrices ${\bf{W}}_1$ and ${\bf{W}}_2$, respectively, where ${\bf{W}}_1$ and ${\bf{W}}_2$ are identically distributed ${\bf{W}} \sim \mathcal{CW}\left(t,\sigma^2{\bf{I}}_s,{\bf{0}}_s\right)$, with underlying Gaussian matrices correlated according to (\ref{eqcorr00}). The exact joint ccdf of $\lambda_s$ and $\varphi_s$ can be expressed as}
\begin{equation}
\label{bivS}
\bar F_{\lambda_s,\varphi_s}\left(\lambda,\varphi\right)=C  \left|{\bf{{\tilde \Upsilon}}}(\lambda,\varphi)\right|,
\end{equation}
\textit{where the entries of the $s\times s$ matrix $\bf{\tilde \Upsilon}$ are given by}
\begin{align}
{\tilde \Upsilon}_{i,j}=2^{(2i-s-t)/2}\cdot\J^{\sigma^{-2} ,\sqrt{2\kappa\varphi} ,\sqrt{2\varepsilon}}_{t-s,s-i,j} \left(\sqrt{\lambda}\right),
\end{align}
\end{theorem}

\begin{IEEEproof}
See Appendix \ref{App1b}.
\end{IEEEproof}

Expressions (\ref{bivL}) and (\ref{bivS}) are given in terms of the IINQ defined in (\ref{eqdef2}). As we show in the following theorem, this IINQ can be expressed in closed-form in terms of a finite sum of Marcum $Q$- functions. 

\begin{theorem}
\label{th01c}
\textit{The IINQ defined in (\ref{eqdef2}) admits a closed-form solution in terms of a finite number of Marcum $Q$-functions given in (\ref{eqprop01}-\ref{eqprop05}).}
\end{theorem}

\begin{IEEEproof}
See Appendix \ref{App5}.
\end{IEEEproof}

Note that in (\ref{eqprop01}), we have defined some auxiliary parameters ${\delta=\alpha+\gamma^2/2}$ and ${\theta=\gamma^2/(\gamma^2+2\alpha)}$, whereas the coefficients $P_{c,l,k}$ and $\omega _{l,k}$ are detailed in Appendix \ref{App5}. In order to obtain this result, two auxiliary integrals (\ref{eqprop02}) and (\ref{eqprop03}) are solved; the proof for these results are given in Appendix \ref{App4}.

\begin{figure*}[t]
\hrulefill
\begin{align}
\label{eqprop01}
\J^{\alpha ,\beta ,\gamma}_{k,c,j} \left(u\right)&= e^ {{ - \tfrac{{\beta ^2 }}{2}\left\{ {1 - \tfrac{{\gamma ^2 }}{{2\alpha }}} \right\}} }\sum\limits_{l = 1}^c {\left\{ {P_{c,l,k} (\beta ^2 )\tfrac{{\gamma ^{l - 1} \beta ^{k + l + 1} }}{{\sqrt {2\delta } ^{2j+\left( {k + l - 1} \right)} }}Q_{2\left( {j - 1} \right)+ (k+l-1)+1,k + l - 1} \left( {\tfrac{{\gamma \beta }}{{\sqrt {2\delta } }},u\sqrt {2\delta } } \right)} \right\}} \nonumber \\ & + \sum\limits_{l = 1}^{c + 1} { {\omega _{l,k} (c)\gamma ^{k + 2\left( {l - 1} \right)} {\mathcal K}^{\alpha ,\beta ,\gamma }_{k+l,j+k+l-1} \left( {u} \right)} },
\end{align}
\hrulefill
%\label{eqprop0y}
\begin{align}
\label{eqprop02}
{\cal K}^{\alpha ,\beta ,\gamma}_{m,n} \left(u\right) &\triangleq \int\limits_u^\infty  {z^{2n - 1} e^{ - \alpha z^2 } Q_{m} \left( {\gamma z,\beta } \right)\:dz}= \frac{\delta^{-n}}{2} \left( {n - 1} \right)!{\mathop{e}\nolimits} ^{ - \frac{{\beta ^2}}{2}}{\mathop{e}\nolimits} ^{-u^2 \delta} \sum\limits_{k = 0}^{n - 1} { \frac{\left( \delta u^2 \right)^k}{{k!}}} \I^{\beta ,\delta ,\theta}_{m,n,k} \left( u \right),\\
\hrulefill
\label{eqprop03}
\I^{{\beta ,\delta,\theta}}_{m,n,k} \left(u\right) &= \left(1 - \theta \right)^{k - n} e^{\tfrac{{\beta ^2 }}{2}} e^{\delta \theta u^2} - \left(1 - \theta \right)^{k - n}\left( {\frac{{\beta ^2 }}{2}} \right)^{k + m - n} \tilde{\Phi} _3 \left( {1,k + m - n + 1;\tfrac{{\beta ^2 }}{2},\tfrac{{\delta \theta u^2 \beta ^2 }}{2}} \right)\nonumber \\&+ \sum\limits_{l = 1}^{n - k} \left(1 - \theta \right)^{k+l-n-1} \left( {\tfrac{{\beta ^2 }}{2}} \right)^{k + m - n + l - 1} \tilde{\Phi} _3 \left( {l,k + m - n + l;\theta \tfrac{{\beta ^2 }}{2},\tfrac{{\delta \theta u^2\beta ^2 }}{2}} \right),
\end{align}
\hrulefill
\begin{align}
\label{eqprop04}
\tilde \Phi _3 \left( {b,c;w,z} \right) &= \left( \frac{z}{w} \right)^{b-1} \sum\limits_{i = 0}^{2(b - 1)}  \frac{\A _i (b,c;z)}{ w^{c - i - 1} z^i } \exp{\left( w + \frac{z}{w} \right)}
Q_{2-c+i} \left( \sqrt{2w}, \sqrt{2\frac{z}{w}} \right) \\
\label{eqprop05}
\A _i (b,c;z)  & = \frac{( - 1)^{b - 1} }{\left( b-1 \right)!} \sum\limits_{k = 0}^{\left\lfloor {i/2} \right\rfloor } {\frac{{( - 1)^k \left( {b - i + k} \right)_{i - k} \left( {c - i - 1 + k} \right)_{i - 2k} }}{{\left( {i - 2k} \right)!k!}} z^{k} }
\end{align}
\hrulefill
\end{figure*}

We observe that the solutions for (\ref{eqprop02}) and (\ref{eqprop03}) are given in terms of the Nuttall $Q$-function, and the regularized confluent hypergeometric function of two variables $\tilde{\Phi}_3(b,c,w,z)$, defined as the classic $\Phi_3$ function in \cite{Gradstein2007} normalized to $\Gamma(c)$. However, for the set of indices in (\ref{eqprop01}), the Nuttall $Q$-function can be computed in terms of the Marcum $Q$-function and the modified Bessel function of the first kind, using the relation defined in \cite{Simon2002} and restated in Appendix \ref{App5} in eq. (\ref{5eqapp02}). With regard to the $\tilde{\Phi}_3$ function, it can also be expressed as a finite number of Marcum $Q$-functions using the relationship recently derived in \cite{Morales2013} and restated in (\ref{eqprop04}). Hence, (\ref{eqprop01}) is given in closed-form in terms of a \textit{finite} number of Marcum $Q$-functions, which are included as built-in functions in most commercial mathematical packages\footnote{Even though the generalized Marcum $Q$-function $Q_m(a,b)$ is mostly used for $m>0$, there exists a simple relation when $m<0$ given in \cite{od2009} as $Q_m(a,b)=1-Q_{1-m}(b,a)$.}.

This closed-form result for the integral (\ref{eqprop01}) is new in the literature to the best of our knowledge, and allows us to directly obtain a closed-form expression for the bivariate distribution of the extreme eigenvalues of a central complex Wishart matrix. These closed-form results for the cdfs have similar form to existing results in the literature for related distributions; specifically, they reduce to the recently calculated expression for the bivariate {Nakagami-$m$} cdf \cite{Lopez2013,Morales2013} when $N_T=m$ and $N_R=1$, as well as to the well-known expression for the bivariate Rayleigh cdf \cite{Schwartz1966} letting $N_T=N_R=1$. For the readers' convenience, a brief description of the $Q$-functions used in this paper is included in Appendix \ref{AppQ}.

\section{Application to MIMO Systems}
\label{Metrics}
%\subsection{MIMO system model}
The dynamic behavior of the random processes of interest, namely the largest and the smallest eigenvalues of a CW matrix, is fully characterized by their joint distribution. In this section, we illustrate how these bivariate cdfs can be used to derive practical metrics for the performance evaluation of MIMO systems.

Let us consider a MIMO system with $N_T$ transmit and $N_R$ receive antennas. In this scenario, the received vector ${{\bf{y}}\in \mathbb{C}^{N_R \times 1}}$ is given by
\begin{equation}
{\bf{y}} = {\bf{H}}{\bf{x}} + {\bf{n}},
\end{equation}
where ${\bf{x}}\in \mathbb{C}^{N_T \times 1}$ is the transmitted vector, ${\bf{n}}\in \mathbb{C}^{N_R \times 1}$ is the noise vector with i.i.d. entries modeled as complex Gaussian RV's $\sim\mathcal{CN}\left(0,1\right)$, and ${\bf{H}}\in \mathbb{C}^{N_R \times N_T}$ represents the Rayleigh fading channel matrix with i.i.d. entries $\sim\mathcal{CN}\left(0,\sigma^2\right)$. 

The MIMO channel can be decomposed into up to $s$ parallel scalar subchannels, where the power gain of the $k^{th}$ eigenchannel depends on the $k^{th}$ eigenvalue of the matrix ${\bf{W}}$. The best and worst achievable performance will be attained using the channels given by the largest and smallest eigenvalues, for which the SNR $\gamma$ is known to be proportional to the magnitude of their respective eigenvalues \cite{Goldsmith2005}, i.e. ${\gamma(t)\propto \lambda}$ and ${\gamma(t+\tau)\propto \varphi}$ according to the notation in Section \ref{sec:PS}. We now define a set of performance metrics that will allow for studying the rate of change of the random processes of interest.

\subsection{Normalized mutual information of SNR values}
\label{discuss}

The joint distributions characterized in this paper incorporate the dynamics of the CW random process through the correlation coefficient $\rho$ of the underlying Gaussian channel matrix, according to the correlation model in eq. (\ref{eqcorr00}). However, the relation between $\rho$ and the correlation coefficient $\rho_i$ of each one of the ordered eigenvalues of the CW matrix is not fully understood. Analytical results for this correlation coefficient are hard to obtain, as they require a two-fold numerical integration over the joint bivariate pdf (or the joint ccdf) of the eigenvalue of interest \cite{McKay2008}.

Simulations in \cite{Kongara2011} show that the rate of change of the largest eigenvalue is much slower than the rate of change of the smaller eigenvalue in a $4\times4$ MIMO setup, i.e. worse eigenchannels decorrelate faster; however, little is known about how the dynamics of MIMO parallel channels change as the number of antennas is increased.

For this reason, we propose an alternative metric to quantify the rate of change of MIMO parallel channels. Instead of deriving the mutual information of $\lambda$ and $\phi$ (i.e. the mutual information of the equivalent SNR $\gamma$ at two different instants), which is still an open problem in the literature \cite{McKay2008,Zhu2009,Matt2010,Chen2012,Li2013}, we study the mutual information of the discrete and identically distributed random variables defined as ${X\triangleq \{\gamma_i(t)<\gamma_{th}\}}$ and ${Y\triangleq \{\gamma_i(t+\tau)<\gamma_{th}\}}$, where $i=1$ and $i=s$ indicate the SNR of the best and worst eigenchannels, respectively.

The mutual information is a measure of the amount of information shared by two random sets $X$ and $Y$. Thus, it is useful to determine how by knowing one of them, the uncertainty about the other one is reduced. In the case of $X$ and $Y$ being two different samples of a random process, this metric gives information about how spacing these two samples affects their independence. The use of normalized mutual information metrics as a measure of similarity is a well investigated subject in statistics \cite{Strehl2003,Vinh2010}. Specifically, we have chosen for our analysis the mutual information metric introduced by \cite{Strehl2003} as
\begin{align}
\label{eqMI01}
\M_{X,Y}\triangleq\overline{I(X,Y)}=\frac{I(X,Y)}{\sqrt{H(X)H(Y)}},
\end{align}

Because they are two samples of a stationary random process, $X$ and $Y$ have the same marginal distribution and therefore $H(X)=H(Y)$. Thus, this metric also reduces to other conventional mutual information metrics \cite{Vinh2010} as
\begin{align}
%\label{eqMI01}
\overline{I(X,Y)}&=\frac{I(X,Y)}{\max{\left\{H(X),H(Y)\right\}}}\nonumber\\&=\frac{2I(X,Y)}{{H(X)+H(Y)}}=\frac{I(X,Y)}{H(X)}=\frac{I(X,Y)}{H(Y)};
\end{align}
which takes values in the range $[0,1]$. Specifically, if ${\rho(\tau)=1}$ then $X$ and $Y$ are the same random variable; thus, ${I(X,X)=H(X)}$ and therefore $\M(\gamma_{th},\tau)=1$. Conversely, letting ${\rho(\tau)=0}$, then $X$ and $Y$ are independent and we have ${I(X,Y)=0\Rightarrow\M(\gamma_{th},\tau)=0}$.

Mutual information metrics are more general than correlation measures, in the sense that they capture dependence other than linear. Besides, the normalized mutual information of discrete random variables can be easily computed, provided that their joint distribution is known. Hence, we use the closed-form expressions for these joint distributions to calculate normalized mutual information of the discrete and identically distributed random variables prevously defined as ${X\triangleq \{\gamma_i(t)<\gamma_{th}\}}$ and ${Y\triangleq \{\gamma_i(t+\tau)<\gamma_{th}\}}$. This metric gives a normalized measure of the rate of change of the random process $\gamma_i(t)$, in terms of the amount of information that is kept between two different samples of the process. Thus, this metric will give information on the similarity between $X$ and $Y$, providing a quantitative mechanism to evaluate the dynamics of the SNR. %In the Appendix \ref{AppMI}, we provide a detailed description of how this metric is calculated.
%\\
%\\
\subsection{Probability of two outage events}
The outage probability is defined as the probability of the instantaneous SNR $\gamma_i$ to be below a certain threshold, i.e. ${P_{out}(\gamma_{th})\triangleq\Pr\{\gamma_i<\gamma_{th}\}}$. This metric does not incorporate any information related with the dynamic variation of $\gamma_i$; however, we can equivalently define the probability of two outage events occurring in two different instants as
\begin{align}
\label{DoubleOP}
{\mathcal{P}_{out}(\gamma_{th},\tau)\triangleq\Pr\{\gamma_i(t)<\gamma_{th},\gamma_i(t+\tau)<\gamma_{th}\}}
\end{align}
where $\tau$ is defined as the separation between two transmissions. Note that this separation has dimensions of time when applied to a time-varying flat-fading MIMO channel, whereas it can take dimensions of frequency when analyzing the equivalent MIMO channel in the frequency domain for MIMO OFDM systems affected by multipath fading.

Since the instantaneous SNR for the eigenchannel associated with a certain eigenvalue is proportional to $\lambda$ \cite{Goldsmith2005}, we can easily see that ${\mathcal{P}_{out}(\gamma_{th},\tau)=F_{\lambda_i,\varphi_i}(\gamma_{th},\gamma_{th})}$. In the limiting case of $\tau\rightarrow\infty$ the process $\gamma_i(t)$ decorrelates and we have $F_{\lambda_i,\varphi_i}(\gamma_{th},\gamma_{th})|_{\tau\rightarrow\infty}=F_{\lambda}(\gamma_{th},\gamma_{th})^2$, whereas for $\tau=0$ we have $F_{\lambda_i,\varphi_i}(\gamma_{th},\gamma_{th})|_{\tau=0}=F_{\lambda_i}(\gamma_{th})$. Intuitively, as the only dependence on $\rho$ in (\ref{bivL}) and (\ref{bivS}) is through $|\rho|^2$, then $\mathcal{P}_{out}(\gamma_{th},\tau)$ is to be bounded above the outage probability and below the square of the outage probability.

\subsection{Level crossing statistics}
The level crossing rate (LCR) is used to determine how often a random process crosses a threshold value. In his seminal work \cite{Rice1944}, Rice introduced a way to compute this metric for continuous processes, using the joint statistics of the random process and its first derivative. This has been the preferred approach for analyzing the LCR in fading channels, as it allows for obtaining compact expressions when the underlying processes are Gaussian.

In some scenarios, the random process of interest is not necessarily continuous. In OFDM systems affected by multipath Rayleigh fading, a discretized equivalent model in the frequency domain is usually assumed, where the channel frequency response at a finite number of subcarriers is defined as a sampled Gaussian random process. In this scenario, the LCR obtained using Rice's approach is only approximated \cite{Kongara2009} and in general overestimates the number of crossings, especially when the condition $\Delta f \cdot \tau<<1$ does not hold, for $\Delta f$ the subcarrier spacing and $\tau$ the rms delay spread. This has an intuitive explanation: when studying the number of crossings of the discretized version of a continuous random process, the possibility of missing a crossing is nonzero, and grows as the sampling period is increased.

An alternative formulation for the LCR analysis of sampled random processes was proposed in \cite{Lopez2012}; interestingly, the LCR can be directly calculated from the bivariate cdf of the random process of interest. Hence, the LCR of the extreme eigenvalues in a MIMO OFDM system can be calculated as
\begin{align}
\label{LCR_D}
N_f(u)=\frac{{F_{\lambda_i}(u)-F_{\lambda_i,\varphi_i}(u,u)}}{\Delta f}.
\end{align}
Equivalently, the average fade duration (AFD) gives information about how long the random process remains below a certain threshold level. This metric can be computed as %using the LCR as
\begin{align}
\label{AFD_D}
A_f(u)=\frac{\Pr\{\lambda_i\leq u\}}{N_{f}(u)}.
\end{align}
In this case, the AFD has dimension of Hertz, and measures the average number of subcarriers undergoing a fade.

\section{Numerical results and discussion}
\label{Applications}
We use the performance metrics defined in Section \ref{Metrics} to analyze the dynamics of the best and worst MIMO parallel channels in different scenarios of interest. In the following figures, markers represent the values obtained by Monte Carlo simulations to check the validity of the theoretical results.

\begin{figure}[t]
	\centering
		\includegraphics[width=\columnwidth]{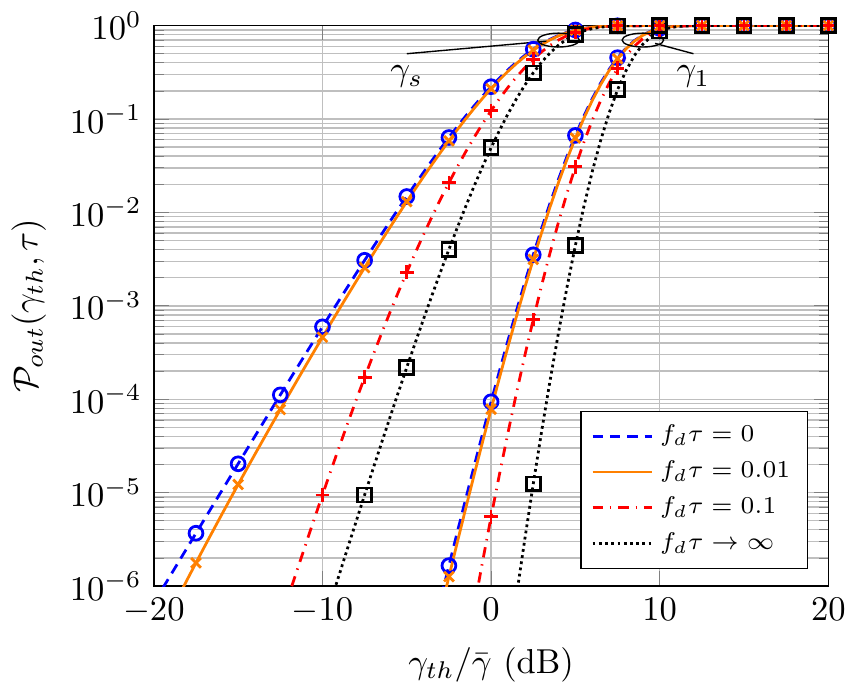}
	\caption{Probability of two outage events vs normalized outage threshold SNR $\gamma_{th}/\bar\gamma$ for different values of $f_d$ and $\tau$.  Parameter values are ${N_T=4}$ and ${N_R=2}$. Markers correspond to Monte Carlo simulations.}
	\label{fig1}
\end{figure}

First, in order to investigate the behavior of the joint distributions derived in this paper, we evaluate the probability of two simultaneous outages in a time-varying MIMO channel. We consider a correlation profile according to Clarke's model, i.e. $\rho=J_0(2\pi f_d \tau)$, where $J_0(\cdot)$ is the Bessel function of the first kind and order zero, $f_d$ is the maximum Doppler frequency and $\tau$ accounts for the time separation between the two outage events. This scenario is illustrated in Fig. \ref{fig1}, where ${\P_{out}(\gamma_{th},\tau)}$ is represented as a function of the threshold SNR $\gamma_{th}$ normalized to the average SNR per branch $\bar \gamma$, for different values of $f_d$ and $\tau$. We assumed a $4\times2$ configuration.

As indicated in Section \ref{Metrics}, the probability of two outage events is bounded by $P_{out}(\gamma_{th})$ and $P_{out}(\gamma_{th})^2$.  We observe how as the separation $\tau$ grows, the two events tend to become independent, and hence ${\P_{out}(\gamma_{th},\tau)\rightarrow{P_{out}(\gamma_{th})^2}}$.  We also note how ${\P_{out}(\gamma_{th},\tau)}$ is confined in a narrower region for the largest eigenvalue, exhibiting a larger variation in the case of the worst channel.

Now, we are interested in understanding how the dynamics of MIMO parallel channels in a time-varying scenario are affected when the number of antennas grows. For this purpose, we use the normalized mutual information metric ${\M(\gamma_{th},\tau)}$ defined in (\ref{eqMI01}). We will pay special attention to the case when $N_R$ is fixed, and $N_T$ is increased. This scenario is of practical interest in the context of large-scale antenna systems, where the number of transmit antennas is much larger than in conventional MIMO systems\footnote{We must note that the results here derived correspond to a Gaussian channel matrix with i.i.d. entries. Even though this assumption does not hold when the number of transmit antennas is very large, it is usually considered as a reference case and in some scenarios it is a reasonably good approximation for the massive linear array case \cite{Larsson2014}.}, and $N_R$ represents the number of single-antenna users. Because of the correlation model in (\ref{eqcorr00}), we consider that all users have a similar mobility characterized by $\rho$.

\begin{figure}[t]
	\centering
		\includegraphics[width=\columnwidth]{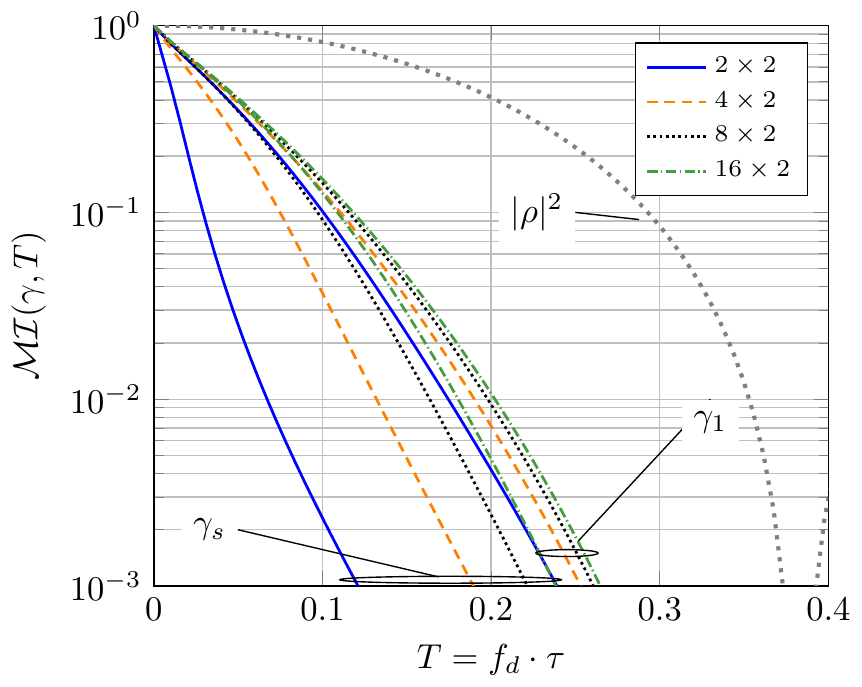}
	\caption{Normalized mutual information $\M(\gamma,T)$ vs $f_d\cdot\tau$ for different numbers of transmit antennas. Parameter values are ${P_{out}(\gamma)=10^{-2}}$, ${N_R=2}$. The dependence on $T$ of $\M(\gamma,T)$ is through $|\rho(T)|^2$ as discussed in \ref{discuss}. The correlation coefficient $\rho$ of the underlying Gaussian MIMO channel is included as a reference.}
	\label{fig1a}
\end{figure}

In Fig. \ref{fig1a}, we represent $\M$ as a function of the parameter $T=f_d\cdot\tau$, for different numbers of transmit antennas $N_T$ and considering $N_R=2$.  This case is very simple, as it considers only two channels; however, it will prove to be very insightful to study the impact of using more transmit antennas in the dynamics of MIMO parallel channels. We assume a value of $\gamma$ that yields an outage probability\footnote{According to the definition of the discrete mutual information metric $\M$, because of the inherent discretization there is a need to define a value of $\gamma$ for which the probabilities are computed. In our case, we decided to chose a value of $\gamma$ that satisfies a certain value of outage probability (OP). Hence, we have computed this value of $\gamma$ by inverting the corresponding cdf (i.e., the OP) for values at which communication systems operate.} of $10^{-2}$. Plots labeled as $\gamma_1$ and $\gamma_s$ correspond to the best and worst eigenchannels, respectively. We have not included markers for the Monte Carlo simulations for the sake of clarity in the plots. However, the coherence between theoretical and simulated results has been checked.

We see how the metric $\M$ for the best eigenchannel is barely affected by using more transmit antennas; in fact, the value of $T$ that achieves a $\M=0.1$ is in the approximate range $[0.10-0.12]$ for the investigated configurations, which corresponds to ${|\rho|^2\approx0.5}$. Conversely, we observe how the dynamic behavior of the worst eigenchannel is dramatically affected by the number of transmit antennas. In this case, the value $\M=0.1$ is attained for a wider set of values of $T$, i.e. $T\approx[0.03-0.12]$. Hence, this indicates that the worst channel decorrelates faster as $N_T$ is reduced. Indeed, the best eigenchannel  takes longer to decorrelate as $N_T$ grows, but this difference is comparatively smaller.

Interestingly, the worst channel rapidly tends to exhibit a similar dynamic behavior as the best eigenchannel as $N_T/N_R$ is increased. In fact, we observe how the best channel in the $2\times2$ case and the worst channel in the $8\times2$ case have similar $\M$. When $16$ transmit antennas are used, the gap between the best and worst channels is small, and the assumption that both channels present a similar dynamic behavior is reasonable.

\begin{figure}[t]
	\centering
		\includegraphics[width=\columnwidth]{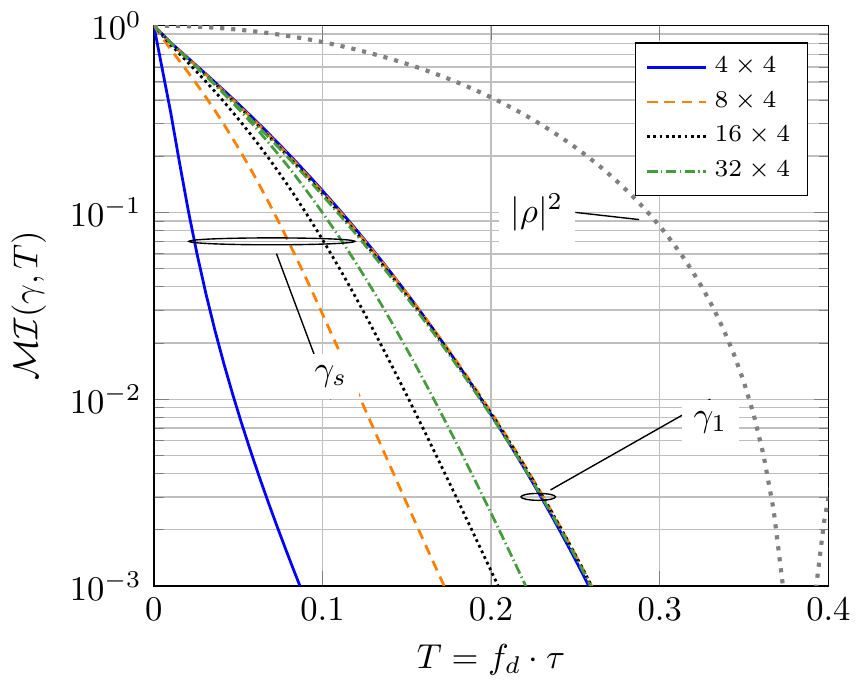}
	\caption{Normalized mutual information $\M(\gamma,T)$ vs $f_d\cdot\tau$ for different numbers of transmit antennas. Parameter values are ${P_{out}(\gamma)=10^{-2}}$, ${N_R=4}$. The dependence on $T$ of $\M(\gamma,T)$ is through $|\rho(T)|^2$ as discussed in \ref{discuss}. The correlation coefficient $\rho$ of the underlying Gaussian MIMO channel is included as a reference.}
	\label{fig1b}
\end{figure}

Fig. \ref{fig1b} considers $N_R=4$, and the same set of parameters as in the previous figure. Now, we observe that the dynamics of the best channel are even more stable, as $\M$ is approximately constant with $N_T$. On the contrary, we see how increasing the number of transmit and receive antennas to $4$ causes the worst channel to have a much faster rate of change. As the number of transmit antennas is increased, we observe again how the worst channel tends to become more stable.

\begin{figure}[t]
	\centering
		\includegraphics[width=\columnwidth]{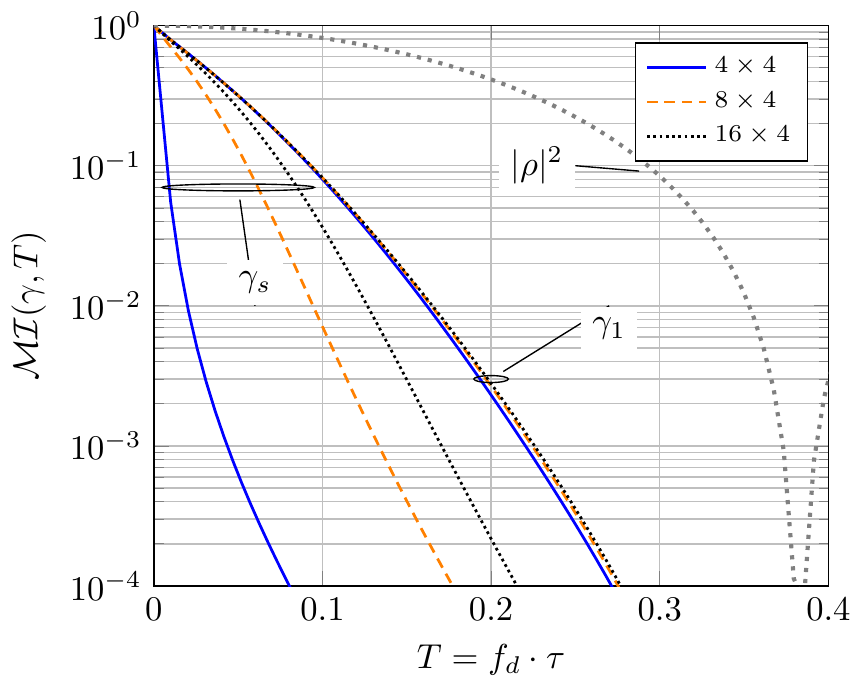}
	\caption{Normalized mutual information $\M(\gamma,T)$ vs $f_d\cdot\tau$ for different numbers of transmit antennas. Parameter values are ${P_{out}(\gamma)=10^{-3}}$, $N_R=4$. The dependence on $T$ of $\M(\gamma,T)$ is through $|\rho(T)|^2$ as discussed in \ref{discuss}. The correlation coefficient $\rho$ of the underlying Gaussian MIMO channel is included as a reference.}
	\label{fig1c}
\end{figure}

In Fig. \ref{fig1c}, we illustrate how the information provided by the metric $\M$ does not strongly depend on the value of $\gamma$. We reproduce the same scenario considered in Fig. \ref{fig1b}, now considering a value of $\gamma$ that yields a OP of $10^{-3}$. Even though the particular values of $\M$ are indeed different, the same conclusions can be inferred.

One of the conclusions extracted in \cite{Larsson2014} stated that for $N_T>10\cdot N_R$, the spread between the best and worst channels is reduced and a stable performance can be ensured even in unfavorable propagation conditions. Here, our results suggest that this performance can be also sustained in time with a similar behavior for all $N_R$ users, assuming that they have a similar mobility.
%\clearpage
Now, we aim to characterize the dynamic behavior of a MIMO OFDM system in terms of the LCR and the AFD. Unlike the conventional Rice approach, the method described in \cite{Lopez2012} allows for computing the level crossing statistics without making any assumption on the differentiability of the autocorrelation function of the random process, and is also valid for discrete correlation models. In this situation, if we consider an exponential multipath profile with rms delay spread $\tau$, then the correlation coefficient of the equivalent channel in the frequency-domain is given by
\begin{align}
\rho=\frac{1}{1-j2\pi\tau\Delta f\cdot |k|},
\end{align}
where $\Delta f$ is the OFDM subcarrier spacing and $k\in\mathbb Z$ is the separation between subcarriers. In the next figures, we evaluate the LCR and the AFD of the extreme eigenchannels for a MIMO OFDM system.

\begin{figure}[t]
	\centering
		\includegraphics[width=.99\columnwidth]{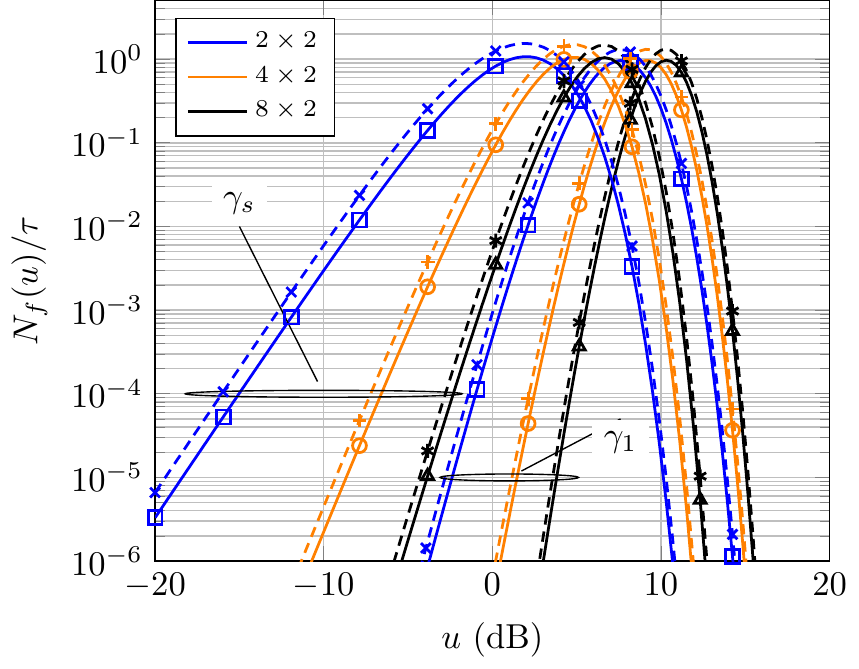}
	\caption{Normalized LCR vs threshold level $u$ for different numbers of transmit antennas and different values of normalized delay spread $\epsilon$. Dashed lines indicate $\epsilon=0.1$, solid lines indicate $\epsilon=0.2$. Parameter values are ${N_R=2}$. Markers correspond to Monte Carlo simulations.}
	\label{fig5}
\end{figure}

\begin{figure}[t]
	\centering
		\includegraphics[width=.99\columnwidth]{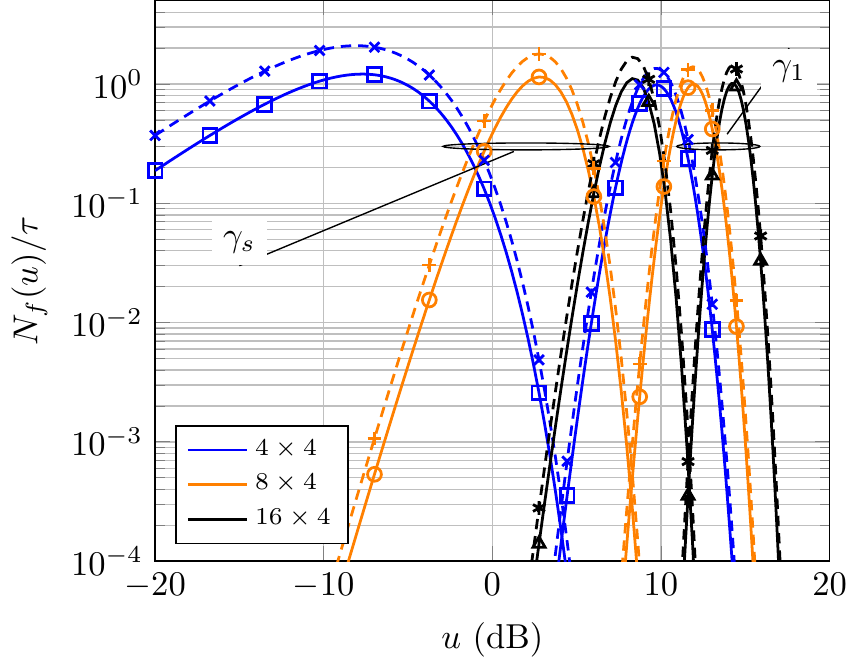}
	\caption{Normalized LCR vs threshold level $u$ for different numbers of transmit antennas and different values of normalized delay spread $\epsilon$. Dashed lines indicate $\epsilon=0.1$, solid lines indicate $\epsilon=0.2$. Parameter values are ${N_R=4}$. Markers correspond to Monte Carlo simulations.}
	\label{fig5b}
\end{figure}

Figs. \ref{fig5} and \ref{fig5b} represent the LCR normalized to $\tau$ for different numbers of transmit antennas and normalized delay spread $\epsilon=\tau\Delta f$, as a function of the normalized threshold value ${u=\gamma_{th}/\bar \gamma}$. Different configurations with two and four receive antennas are considered, and Monte Carlo simulations are included with markers. When the number of transmit antennas is increased, we observe how the value of $u$ for which the maximum number of crossings occurs also grows; however, this effect is more noticeable for the worst channel. Note also that LCR curves tend to present less variance as more transmit antennas are used, but again this trend is especially noteworthy for the smallest eigenvalue. Hence, this also confirms that the rate of change of the worst eigenchannel becomes more stable as $N_T$ is increased. 

We see how the LCR of the largest eigenvalue when using four receive antennas has a similar behavior as in the two receive antennas counterpart, when increasing the number of Tx antennas. The shape of the LCR curves for the largest eigenvalue is narrower in the 4-Rxantenna case, and increasing the number of transmit antennas shifts figures to the right. For the smallest eigenvalue, we observe how a larger number of crossings occur for low values of the threshold level $u$. We also see that increasing the number of Tx antennas has a dramatic effect on the shape of the LCR curves, which are very spread out when $N_T=N_R$, and tend to have a similar shape as the largest eigenvalue LCR when $N_T$ is sufficiently large. We note how the shapes of the LCR curves for the best channel in the $4\times4$ case, and the worst channel in the $16\times4$ case, are very similar, being only shifted around $1.2$ dB.

\begin{figure}[t]
	\centering
		\includegraphics[width=.99\columnwidth]{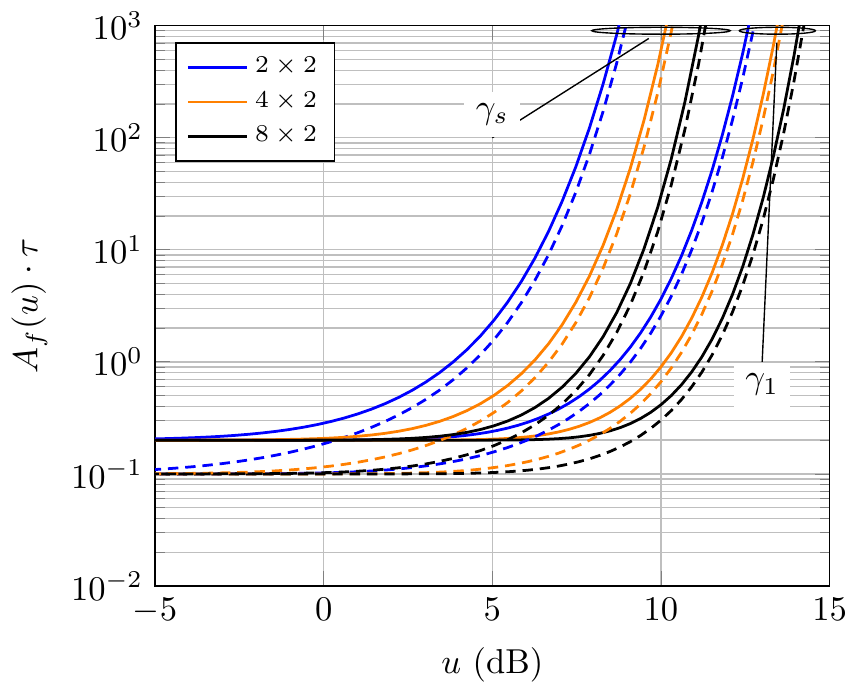}
	\caption{Normalized AFD vs threshold level $u$ for different numbers of transmit antennas and different values of normalized delay spread $\epsilon$. Dashed lines indicate $\epsilon=0.1$, solid lines indicate $\epsilon=0.2$. Parameter values are ${N_R=2}$.}
	\label{fig6}
\end{figure}

\begin{figure}[t]
	\centering
		\includegraphics[width=.99\columnwidth]{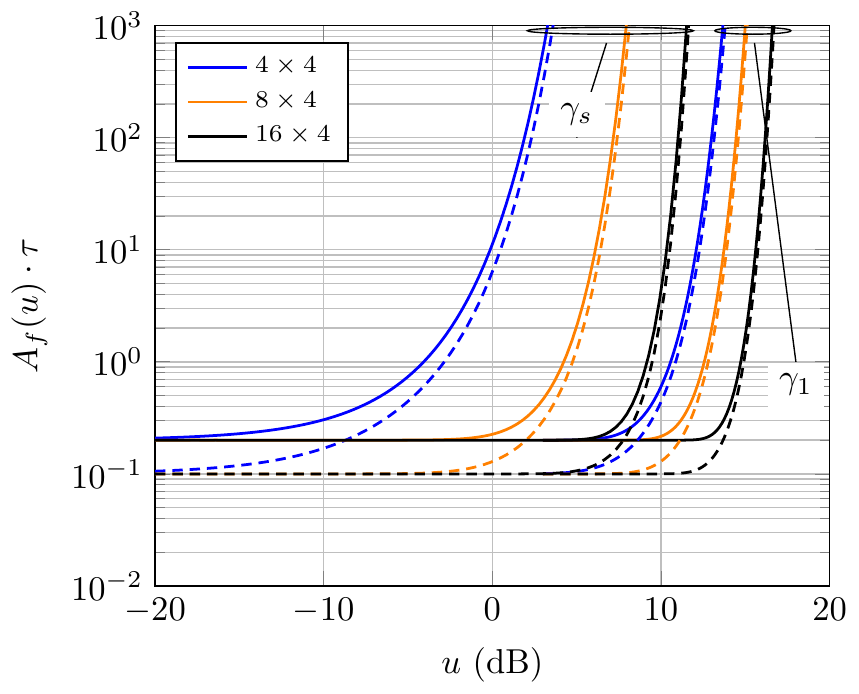}
	\caption{Normalized AFD vs threshold level $u$ for different numbers of transmit antennas and different values of normalized delay spread $\epsilon$. Dashed lines indicate $\epsilon=0.1$, solid lines indicate $\epsilon=0.2$. Parameter values are ${N_R=4}$.}
	\label{fig6b}
\end{figure}

Figs. \ref{fig6} and \ref{fig6b} represent the AFD normalized with $\tau$ for different numbers of transmit antennas and delay spread, as a function of the threshold value $u$. Again, configurations with two and four receive antennas are considered. For a fixed value of the threshold level $u$ used to declare a fade, we observe how the AFD is reduced when more transmit antennas are considered. This is consistent with the fact that a better SNR is expected when more antennas are used. We also note how the normalized AFD tends asymptotically to $\epsilon$ as ${u\rightarrow\infty}$, i.e. the AFD tends asymptotically to $\Delta f$. This behavior has an intuitive explanation based on the discrete definition of the AFD: in the event that a fade is declared on a given subcarrier, the minimum duration of a fade is exactly the distance to the closest subcarrier. We also see that, in general, the AFD figures are now more abrupt when 4 receive antennas are used, compared to the case of 2 receive antennas.

%\clearpage
\section{Conclusions}
We obtained exact closed-form expressions for the joint cdfs of the extreme eigenvalues of complex central Wishart matrices. Despite the inherent complexity of the analyzed distributions, the results here derived have a similar form as other related distributions in the context of communication theory, and can be computed with well-know special functions included in commercial mathematical packages.

We have used these results to characterize the dynamics of the parallel channels associated with MIMO transmission. The analytical results here obtained for the best and worst channels are useful to study the evolution of the SNRs in MIMO Rayleigh fading channels. 
We illustrated how different performance metrics can be easily evaluated: In addition to well-known metrics such as the LCR, the AFD or the probability of two outage events, we used a \textit{normalized mutual information} metric to characterize the rate of change of MIMO parallel channels directly from the derived joint cdf. Other performance metrics like the transition probabilities for a first-order Markov chain model \cite{Zorzi1998} can be calculated from these distributions.

We have observed that when the number of transmit and receive antennas is similar, the worst channel has a much faster variation than the best channel. While the dynamics of the latter are barely affected by using more transmit antennas, we notice that the worst channel tends to have a more stable behavior as $N_T$ is increased. This suggests that the user channel variation in massive MIMO systems would be similar for all users with the same mobility.

The derivation of asymptotic limit results for large $N_T$ and $N_R$ would be of general interest in random matrix theory. Unfortunately, even though our results for the bivariate distributions analyzed are given in closed-form for the first time in the literature, they are still quite involved and not easily generalizable to the asymptotic regime. In fact, a different approach than the one taken herein might be required for this new analysis. Therefore, characterizing the dynamics of the extreme eigenvalues in the limit of large $N_T$ and $N_R$ is left as a topic for future research.

\label{Conclusion}

\section*{Acknowledgments}
The authors would like to thank Dr. David Morales-Jimenez for insightful discussions and for his assistance on elaborating Fig. \ref{figContour01}.
\appendices
%\clearpage

\section{Joint bivariate cdf of the largest eigenvalue}
\label{App1}
The random matrix ${\bf{W_2}}|{\bf{W_1}}$ follows a non-central CW distribution according to (\ref{CWdist01}). Hence, the cdf of the largest eigenvalue of ${\bf{W_2}}|{\bf{W_1}}$ is given by \cite{Jin2008}
\begin{align}
\label{eqapp01}
F_{\varphi _1 |\Lambda} &(\varphi ,\lambda _1,...,\lambda _s ) = \left| {\bf{C}} \right|\left|{\bf{G}}\left(0,\Lambda\right)-{\bf{G}}\left(\varphi,\Lambda\right)\right|
\end{align}
where the entries of matrix ${\bf{G}}\left(\varphi,\Lambda\right)$ are given by
\begin{align}
\label{eqapp02}
G_{i,j}\left(\varphi,\lambda_j\right) = \bar \lambda _j^{\frac{{s - t}}{2}} e^{\bar \lambda _j } 2^{\frac{{2i - s - t}}{2}} Q_{a,b} \left( {\sqrt {2\bar {\lambda }_j}  ,\sqrt {2\bar \varphi} } \right),
\end{align}
for $\{i,j\}=1\ldots s$, ${a=s + t - 2i + 1}$, ${b=t - s}$, $Q_{m,n}(\cdot,\cdot)$ is the Nuttall $Q$-function, the determinant $\left| {\bf{C}} \right|$ is given by
\begin{align}
\label{eqapp03}
\left| {\bf{C}} \right| = {{e^{ - \sum\limits_{j = 1}^s {\bar \lambda _j } } }}/{{\prod\limits_{i < j}^s {\left( {\bar \lambda _i  - \bar \lambda _j } \right)} }}
\end{align}
and $\bar \lambda _j$ and $\bar \varphi$ are defined as scaled versions of $\lambda_j$ and $\varphi$ respectively
\begin{align}
\label{eqapp04}
\bar \lambda _j & \buildrel \Delta \over = \frac{{\rho ^2 }}{{\left( {1 - \rho ^2 } \right)\sigma ^2 }}\lambda _j {\rm{ = }}\varepsilon \lambda _j,\\ \bar \varphi &\triangleq \frac{{1 }}{{\left( {1 - \rho ^2 } \right)\sigma ^2 }} \varphi = \kappa \varphi.
\end{align}
For convenience of calculation, we re-express
\begin{align}
\label{eqapp05}
\left| {\bf{C}} \right| = c_s \left| {\bf{E}} \right|/\left| {\bf{V}} \right|,
\end{align}
where ${\bf{E}}$ is a diagonal matrix whose entries are given by $E_{j,j}  = {{e^{ - \bar \lambda _j } }}$, ${\bf{V}}$ is a Vandermonde matrix with entries $V_{i,j}  = \lambda _i^{j - 1}$, and
\begin{align}
\label{eqcs}
c_s=\frac{(-1)^{s(s-1)/2}}{\prod_{i=0}^{s-1}{\varepsilon^i}}.
\end{align}
Note that both $\bf{E}$ and $\bf{V}$ matrices depend on the eigenvalues ${{\bf{\lambda}}=\left[\lambda_1\ldots \lambda_s\right]^T}$, although this dependence is omitted for the sake of compactness. Analogously, we split (\ref{eqapp01}) into three terms as
\begin{align}
\label{eqapp08}
{\bf{G}}\left(0,\Lambda\right)-{\bf{G}}\left(\varphi,\Lambda\right)  = {\bf{A Y B}},
\end{align}
where
\begin{align}
\label{eqapp09}
B_{i,j} (\varphi) = 2^{\frac{{2i - s - t}}{2}} \left\{ {Q_{a,b} \left( {\sqrt {2\bar \lambda _j } ,0} \right) - Q_{a,b} \left( {\sqrt {2\bar \lambda _j } ,\sqrt {2\bar\varphi} } \right)} \right\},
\end{align}
and ${\bf{A}},$${\bf{Y}}$ are diagonal matrices whose entries are given by $A_{{j,j}} = {\varepsilon ^{\frac{{\left( {s - t} \right)}}{2}} }e^{\bar \lambda _j }$, and $Y_{{j,j}}  = \lambda _j^{\frac{{s - t}}{2}}$. Again, the dependence on $\bf{\lambda}$ is omitted in these matrices. With these definitions, we can express (\ref{eqapp01}) as
\begin{align}
\label{eqapp12}
F_{\varphi_1 |\Lambda} &(\varphi ,\lambda _1,...,\lambda _s ) = c_s \varepsilon ^{\frac{{s\left( {s - t} \right)}}{2}} \left| {\bf{Y}} \right|\left| {{\bf{B}}(\varphi)} \right|/\left| {\bf{V}} \right|.
\end{align}

In order to obtain the joint cdf $F_{\lambda _1 ,\varphi _1 } (\lambda ,\varphi )$ of the largest eigenvalues $\{\lambda_1,\varphi_1\}$, we average (\ref{eqapp12}) using the joint pdf of the eigenvalues $\{\lambda_1,\ldots,\lambda_s\}$ given by \cite{Edelman1989}
\begin{align}
\label{eqapp13}
f_{\Lambda} (\lambda _1,...,\lambda _s ) = K\left| {\bf{Z}} \right|\left| {\bf{V}} \right|^2,\,\,\,\,\,\, \lambda _1  > \lambda _2  > ... > \lambda _s  > 0,
\end{align}
where ${\bf{Z}}$ is a diagonal matrix with entries given by
\begin{align}
\label{eqapp14}
Z_{j,j}={\lambda _i ^{t - s} e^{ - \lambda _i /\sigma ^2 }},
\end{align}
${\bf{V}}$ is the Vandermonde matrix defined in (\ref{eqapp05}), and
\begin{align}
\label{eqapp15}
K = \frac{{\left( {\sigma ^2 } \right)^{ - t\cdot s} }}{{\prod\limits_{i = 1}^s {\Gamma \left( {s - i + 1} \right)\Gamma \left( {t - i + 1} \right)} }},
\end{align}
is a normalization factor. Thus, the joint cdf can be expressed as
\begin{align}
F_{\lambda _1 ,\varphi _1 }& (\lambda ,\varphi ) = Kc_s \varepsilon ^{\frac{{s\left( {s - t} \right)}}{2}} \underbrace {\mathop {\int { \ldots \int {} } }\limits_{\lambda  > \lambda _1  > ...\lambda _s  > 0} }_{s - fold}\left| {\bf{\Theta }} \right|\left| {{\bf{B}}(\varphi )} \right|d\lambda _1 ...d\lambda _s, \label{eqapp16l}
\end{align}
where ${\bf{\Theta }}={\bf{YVZ}}$. Using \cite[eq. 51]{Chiani2003} we can express the $s-$fold integral of a determinant, as
\begin{align}
F_{\lambda _1 ,\varphi _1 } (\lambda ,\varphi ) = Kc_s  \varepsilon ^{\frac{{s\left( {s - t} \right)}}{2}} \left| {\int\limits_0^\lambda  {{\bf{\Psi }}(x)} dx} \right|,
 \label{eqapp18}
\end{align}
where the entries of the $s\times s$ matrix $\bf\Psi$ are given by
\begin{align}
\Psi_{i,j}=&2^{(2i - t - s)/2} e^{ - \frac{x}{{\sigma ^2 }}} x^{j + \frac{{t - s}}{2} - 1}\nonumber\\\times&\left\{ {Q_{\iota ,t - s} \left( {\sqrt {2\varepsilon x} ,0} \right) - Q_{\iota ,t - s} \left( {\sqrt {2\varepsilon x} ,\sqrt {2\kappa \varphi } } \right)} \right\},
\end{align}
and $\iota=2(s-i)+(t-s)+1$. Using a change of variables $x=z^2$, we can express the integral (\ref{eqapp18}) as
\begin{align}
F_{\lambda _1 ,\varphi _1 } (\lambda ,\varphi ) = C  \left|{\bf{\Upsilon}}\right|,
\label{eqapp20}
\end{align}
where
\begin{align}
\label{eqC}
C =2^s \sigma ^{ - 2ts} ( - 1)^{\frac{{s(s - 1)}}{2}} \left\{ {\tfrac{{\rho ^2 }}{{(1 - \rho ^2 )\sigma ^2 }}} \right\}^{\frac{{s(s - t)}}{2}} \prod\limits_{i = 1}^s {\tfrac{{\left\{ {\frac{{\rho ^2 }}{{(1 - \rho ^2 )\sigma ^2 }}} \right\}^{1 - i} }}{{\left( {s - i} \right)!\left( {t - i} \right)!}}},
\end{align}
and the entries of the $s\times s$ matrix $\bf{\Upsilon}$ are given by
\begin{align}
\Upsilon_{i,j}=&2^{(2i-s-t)/2}\left\{\J^{\sigma^{-2} ,0 ,\sqrt{2\varepsilon}}_{t-s,s-i,j} \left(0\right)-\J^{\sigma^{-2} ,\sqrt{2\kappa\varphi} ,\sqrt{2\varepsilon}}_{t-s,s-i,j} \left(0\right)\nonumber\right.\\ &\left.-\J^{\sigma^{-2} ,0 ,\sqrt{2\varepsilon}}_{t-s,s-i,j} \left(\sqrt{\lambda}\right)+
\J^{\sigma^{-2} ,\sqrt{2\kappa\varphi} ,\sqrt{2\varepsilon}}_{t-s,s-i,j} \left(\sqrt{\lambda}\right)\right\},
\end{align}
using the definition given in (\ref{eqprop01}) for the IINQ function. Since a closed-form expression for $\J^{\alpha ,\beta ,\gamma}_{k,c,j} \left(y\right)$ is given in (\ref{eqprop01}), this yields a closed-form expression for the joint cdf ${F_{\lambda _1 ,\varphi _1 } (\lambda ,\varphi )}$ in (\ref{eqapp18}).
%
%\clearpage

\section{Joint bivariate ccdf of the smallest eigenvalue}
\label{App1b}
This proof is similar to the one detailed in the previous appendix. The random matrix ${\bf{W_2}}|{\bf{W_1}}$ follows a non-central CW distribution according to (\ref{CWdist01}). Hence, the ccdf of the smallest eigenvalue of ${\bf{W_2}}|{\bf{W_1}}$ is given by \cite{Jin2008}
\begin{align}
\label{eqapp01s}
\bar F_{\varphi _s |\Lambda} (\varphi ,\lambda _1,...,\lambda _s ) = \left| {\bf{C}} \right|\left|{\bf{G}}\left(\varphi,\Lambda\right)\right|
\end{align}
where the entries of matrix ${\bf{G}}\left(\varphi,\Lambda\right)$ are given in (\ref{eqapp02}), and the determinant $\left| {\bf{C}} \right|$ is detailed in (\ref{eqapp03}). Following a similar reasoning as in Appendix \ref{App1}, we can re-express (\ref{eqapp01s}) as
\begin{align}
\label{eqapp12s}
\bar F_{\varphi _s |\Lambda} &(\varphi ,\lambda _1,...,\lambda _s ) = c_s \varepsilon ^{\frac{{s\left( {s - t} \right)}}{2}} \left| {\bf{Y}} \right|\left| {{\bf{\tilde B}}(\varphi)} \right|/\left| {\bf{V}} \right|,
\end{align}
where
\begin{align}
\label{eqapp09s}
{\tilde B}_{i,j} (\varphi) = 2^{\frac{{2i - s - t}}{2}}  { Q_{a,b} \left( {\sqrt {2\bar \lambda _j } ,\sqrt {2\bar\varphi} } \right)},
\end{align}
and the rest of parameters were defined in the previous proof. In order to obtain the joint ccdf $F_{\lambda _s ,\varphi _s } (\lambda ,\varphi )$ of the smallest eigenvalues $\{\lambda_s,\varphi_s\}$, we average (\ref{eqapp12s}) using the joint pdf of the eigenvalues $\{\lambda_1,\ldots,\lambda_s\}$ given in (\ref{eqapp13}). After some manipulations, the joint ccdf can be expressed as
\begin{align}
F_{\lambda _s ,\varphi _s }& (\lambda ,\varphi ) = Kc_s \varepsilon ^{\frac{{s\left( {s - t} \right)}}{2}} \underbrace {\mathop {\int { \ldots \int {} } }\limits_{\infty>\lambda  > \lambda _1  > ...\lambda _s } }_{s - fold}\left| {\bf{\Theta }} \right|\left| {{\bf{\tilde B}}(\varphi )} \right|d\lambda _1 ...d\lambda _s, \label{eqapp16s}
\end{align}
Using\cite[eq. 51]{Chiani2003} we can express the $s-$fold integral of a determinant, as
\begin{align}
F_{\lambda _s ,\varphi _s } (\lambda ,\varphi ) = Kc_s  \varepsilon ^{\frac{{s\left( {s - t} \right)}}{2}} \left| {\int\limits_\lambda^\infty  {{\bf{\tilde\Psi }}(x)} dx} \right|,
 \label{eqapp18s}
\end{align}
where the entries of the $s\times s$ matrix $\bf{\tilde\Psi}$ are given by
\begin{align}
{\tilde \Psi}_{i,j}=&2^{(2i - t - s)/2} e^{ - \frac{x}{{\sigma ^2 }}} x^{j + \frac{{t - s}}{2} - 1} { Q_{\iota ,t - s} \left( {\sqrt {2\varepsilon x} ,\sqrt {2\kappa \varphi } } \right)},
\end{align}
and $\iota=2(s-i)+(t-s)+1$. Using a change of variables $x=z^2$, we can express the integral (\ref{eqapp18s}) as
\begin{align}
F_{\lambda _s ,\varphi _s } (\lambda ,\varphi ) = C  \left|{\bf{\tilde\Upsilon}}\right|,
\label{eqapp20s}
\end{align}
where $C$ is given in (\ref{eqC}) and the entries of the $s\times s$ matrix $\bf{\tilde \Upsilon}$ are given by
\begin{align}
{\tilde\Upsilon}_{i,j}=2^{(2i-s-t)/2}\J^{\sigma^{-2} ,\sqrt{2\kappa\varphi} ,\sqrt{2\varepsilon}}_{t-s,s-i,j} \left(\sqrt{\lambda}\right),
\end{align}
using the definition given in (\ref{eqprop01}) for the IINQ function. Since a closed-form expression for $\J^{\alpha ,\beta ,\gamma}_{k,c,j} \left(y\right)$ is given in (\ref{eqprop01}), this yields a closed-form expression for the joint ccdf ${F_{\lambda _s ,\varphi _s } (\lambda ,\varphi )}$ in (\ref{eqapp18s}).

\section{Appendix: Solution for $\J_{k,c,j}^{\alpha ,\beta ,\gamma} \left(u\right)$}
\label{App5}
Let us consider the incomplete integral defined in (\ref{eqprop01}) as
\begin{align}
\label{5eqapp01}
\J_{k,c,j}^{\alpha ,\beta ,\gamma} \left(u\right) = \int\limits_y^\infty  {z^{2j + k - 1} e^{ - \alpha z^2 } Q_{2c + k + 1,k} \left( {\gamma z,\beta } \right)\:dz},
\end{align}
where $k,c,j\in \mathbb{N}$ and $\alpha ,\beta ,\gamma, u  \in \mathbb{R}^+$. The Nuttall $Q_{m,n}$ function can be expressed in terms of a finite sum of Marcum $Q_k(\cdot,\cdot)$ and modified Bessel functions of the first kind $I_n(\cdot)$, when the sum of indices $m+n$ is an odd number \cite[eq. 8]{Simon2002} as
\begin{align}
\label{5eqapp02}
& Q_{2c + k + 1,k} \left( {\gamma z,\beta } \right) = \sum\limits_{l = 1}^{c + 1} { {{\omega _{l,k} (c)}(\gamma z)^{k + 2\left( {l - 1} \right)} Q_{k + l} \left( {\gamma z,\beta } \right)}  }\nonumber\\+ & e^{ - \frac{{\left( {\gamma z} \right)^2  + \beta ^2 }}{2}} \sum\limits_{l = 1}^c {\left\{ {P_{c,l,k} (\beta ^2 )\left( {\gamma z} \right)^{l - 1} \beta ^{k + l + 1} I_{k + l - 1} \left( {\gamma z\beta } \right)} \right\}},
\end{align}
where
\begin{align}
\label{5eqapp03}
\omega _{l,k} (c) &= 2^{c - l + 1} \frac{{c!}}{{(l - 1)!}}\left( {\begin{array}{*{20}c}
   {c + k}  \\
   {c - l + 1}  \\
\end{array}} \right),\\
\label{5eqapp04}
P_{c,l,k} (\beta ^2 ) &= \sum\limits_{r = 0}^{c - l} {2^{c - l - r} \frac{{\left( {c - 1 - r} \right)!}}{{(l - 1)!}}\left( {\begin{array}{*{20}c}
   {c + k}  \\
   {c - l - r}  \\
\end{array}} \right)\beta ^{2r} }.
\end{align}

Substituting (\ref{5eqapp02}) into (\ref{5eqapp01}), we split the IINQ into a finite sum of Nuttall $Q$-functions and incomplete integrals of the Marcum $Q$-function as in (\ref{eqprop01}). The integral denoted as ${\cal K}^{\alpha ,\beta ,\gamma}_{m,n}(u)$ in (\ref{eqprop02}) can be seen as a generalization of that given in \cite{Lopez2013}; the solution for this integral is given in Appendix \ref{App4}.

\section{Appendix: Solution for ${\cal K}_{m,n} ^{\alpha ,\beta ,\gamma}(u)$ }
\label{App4}

\begin{figure}[t]
\begin{center}
\includegraphics[width=\columnwidth]{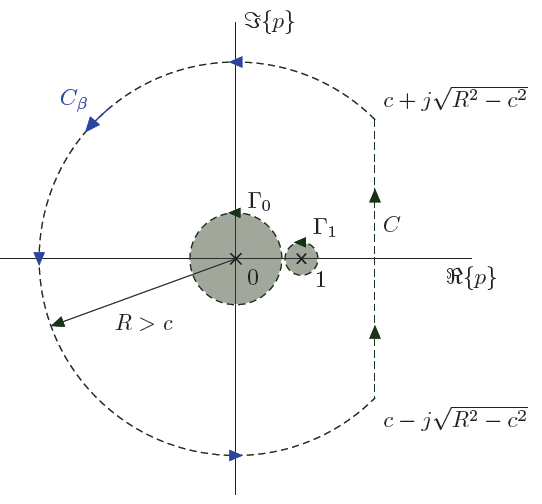}
\caption{Contour integration for integral ${I}_{m,n,k}$.}
\label{figContour01}
\end{center}
\end{figure}

We aim to find a closed-form expression for the integral
\begin{align}
\label{4eqapp01}
{\cal K}_{m,n} ^{\alpha ,\beta ,\gamma}(u)= \int_u^\infty  {x^{2n - 1} e^  { - \alpha x^2 }Q_m \left( {\gamma x,\beta } \right)dx} ,
\end{align}
where $n , m \in \mathbb{N}$ and $\alpha ,\beta ,\gamma, u  \in \mathbb{R}^+$, which is a generalization of that given in \cite{Lopez2012}. The Marcum $Q$-function can be expressed in terms of a contour integral as \cite{Proakis2001}
\begin{align}
\label{4eqapp02}
Q_m \left( {\gamma x,\beta } \right) = e^ { - \frac{{\gamma ^2 x^2  + \beta ^2 }}{2}}\oint_\Gamma  {\frac{1}{{p^m }}\frac{1}{{1 - p}}e^ {\frac{{\gamma ^2 x^2 }}{{2p}} + \frac{{\beta ^2 p}}{2}}dp},
\end{align}
where $\Gamma$ is a circular contour of radius less than unity enclosing the origin. Thus, we can express
\begin{align}
\label{4eqapp03}
{\cal K}_{m,n} ^{\alpha ,\beta ,\gamma}(u)& = \int_u^\infty  x^{2n - 1} e^{ - \alpha x^2 } \nonumber\\ &\times \left\{ {e^ { - \frac{{\gamma ^2 x^2  + \beta ^2 }}{2}} \oint_\Gamma  {\frac{1}{{p^m }}\frac{1}{{1 - p}}e^ {\frac{{\gamma ^2 x^2 }}{{2p}} + \frac{{\beta ^2 p}}{2}} dp} }\right\} dx.
\end{align}
After some manipulations, and letting $\delta  = \alpha  + \gamma ^2 /2$ we have
\begin{align}
\label{4eqapp04}
{\cal K}_{m,n} ^{\alpha ,\beta ,\gamma}&(u) =\\ &\nonumber{\mathop{e}\nolimits} ^{ - \frac{{\beta ^2 }}{2}} \int_u^\infty  {x^{2n - 1} {\mathop{e}\nolimits} ^{ - \delta x^2 }}\left\{ {\oint_\Gamma  {\frac{1}{{p^m }}\frac{1}{{1 - p}}{\mathop{e}\nolimits} ^{\frac{{\gamma ^2 x^2 }}{{2p}}} {\mathop{e}\nolimits} ^{\frac{{\beta ^2 p}}{2}} dp} } \right\}dx,
\end{align}
and changing the integration order
\begin{align}
\label{4eqapp05}
{\cal K}_{m,n} ^{\alpha ,\beta ,\gamma}&(u) = \nonumber\\&{\mathop{e}\nolimits} ^{ - \frac{{\beta ^2 }}{2}} \oint_\Gamma  {\frac{1}{{p^m }}\frac{1}{{1 - p}}{\mathop{e}\nolimits} ^{\frac{{\beta ^2 p}}{2}} \left\{ {\int_u^\infty  {x^{2n - 1} {\mathop{e}\nolimits} ^{ - \delta x^2 } {\mathop{e}\nolimits} ^{\frac{{\gamma ^2 x^2 }}{{2p}}} dx} } \right\}dp}.
\end{align}
Let $\varepsilon =2\delta - \gamma ^2 /p$; the inner integral in (\ref{4eqapp05}) is given by
\begin{align}
\label{4eqapp06}
\int_u^\infty  {x^{2n - 1} {\mathop{e}\nolimits} ^{ - \frac{\varepsilon x^2}{2} } dx}  = \varepsilon ^{ - n} 2^{n - 1} \Gamma \left( {n,\frac{{u^2 \varepsilon }}{2}} \right)
\end{align}
where $\Gamma(n,w)$ is the upper incomplete Gamma function. Since $n$ is a positive integer, we can use the following relationship
\begin{align}
\label{4eqapp07}
\Gamma \left( {n,w} \right) = \left( {n - 1} \right)!\exp \left( { - w} \right)\sum\limits_{k = 0}^{n - 1} {\frac{{w^k }}{{k!}}},
\end{align}
to express (\ref{4eqapp05}) as
\begin{align}
\label{4eqapp08}
{\cal K}_{m,n} ^{\alpha ,\beta ,\gamma}&(u)= {\mathop{e}\nolimits} ^{ - \frac{{\beta ^2 }}{2}} \oint_\Gamma  \frac{1}{{p^m }}\frac{1}{{1 - p}}{\mathop{e}\nolimits} ^{\frac{{\beta ^2 p}}{2}}\\ &\times \left\{ {\varepsilon ^{ - n} 2^{n - 1} \left( {n - 1} \right)!e^{ - \frac{{u^2 \varepsilon }}{2}} \sum\limits_{k = 0}^{n - 1} {\left( {\frac{{u^2 \varepsilon }}{2}} \right)^k \frac{1}{{k!}}} } \right\}dp,\nonumber
\end{align}
which can be conveniently rearranged as
\begin{align}
\label{4eqapp09}
{\cal K}_{m,n} ^{\alpha ,\beta ,\gamma}&(u) = 2^{n - 1} \left( {n - 1} \right)!{\mathop{e}\nolimits} ^{ - \frac{{\beta ^2 }}{2}}\\ &\times\sum\limits_{k = 0}^{n - 1} {\left( {\frac{{u^2 }}{2}} \right)^k \frac{1}{{k!}}} \oint_\Gamma  {\frac{{\varepsilon ^{ - (n - k)} }}{{p^m }}\frac{1}{{1 - p}}{\mathop{e}\nolimits} ^{\frac{{\beta ^2 p}}{2}} e^{ - \frac{{u^2 \varepsilon }}{2}} dp}.\nonumber
\end{align}
If we define $\theta  = \gamma ^2 /2\delta$, we have
\begin{align}
\label{4eqapp10}
{\cal K}_{m,n} ^{\alpha ,\beta ,\gamma}&(u)= \frac{\delta^{-n}}{2}  \left( {n - 1} \right)!{\mathop{e}\nolimits} ^{ - \frac{{\beta ^2 }}{2}} e^{ - u^2 \delta} \\ &\times\sum\limits_{k = 0}^{n - 1} { \frac{\left( u^2 \delta \right)^k}{{k!}}} \oint_\Gamma  {\frac{{\left( {1 - \theta p^{ - 1} } \right)^{ - (n - k)} }}{{p^m }}\frac{1}{{1 - p}}{\mathop{e}\nolimits} ^{\frac{{\beta ^2 p}}{2}} e^{\frac{{\delta \theta u^2 }}{{p}}} dp}\nonumber.
\end{align}
After some algebra, we obtain
\begin{align}
\label{4eqapp11}
{\cal K}_{m,n} ^{\alpha ,\beta ,\gamma}(u)= \frac{\delta^{-n}}{2} \left( {n - 1} \right)!{\mathop{e}\nolimits} ^{ - \frac{{\beta ^2 }}{2}} e^{ - u^2 \delta} \sum\limits_{k = 0}^{n - 1} { \frac{\left( u^2 \delta \right)^k}{{k!}}} I_{m,n,k}^{\beta ,\delta ,\theta} \left(u\right),
\end{align}
where
\begin{align}
\label{4eqapp12}
I_{m,n,k}^{\beta ,\delta ,\theta} \left( u \right) \buildrel \Delta \over =  - \oint_\Gamma  {\frac{1}{{p^{k - (n - m)} }}\frac{1}{{p - 1}}\frac{1}{{\left( {p - \theta } \right)^{n - k} }}{\mathop{e}\nolimits} ^{\frac{{\beta ^2 p}}{2}} e^{\frac{{\delta \theta u^2 }}{{p}}} dp}.
\end{align}
Let us define the function $G(p)$ as
\begin{align}
\label{4eqapp13}
G(p) \buildrel \Delta \over = \frac{1}{{p^{k - (n - m)} }}\frac{1}{{p - 1}}\frac{1}{{\left( {p - \theta } \right)^{n - k} }}e^{\frac{{\delta \theta u^2 }}{{p}}}.
\end{align}
Thus, we have
\begin{align}
\label{4eqapp14}
I_{m,n,k}^{\beta ,\delta ,\theta} \left( u \right)  =  - \oint_\Gamma  {G(p){\mathop{e}\nolimits} ^{\frac{{\beta ^2 p}}{2}} dp}.
\end{align}
Interestingly, the integrand is in the form of an inverse Laplace transform. Hence, we aim to find a connection between this integral defined in the contour $\Gamma$ and the general Bromwich integral
\begin{align}
\label{4eqapp15}
g(t) = \frac{1}{{2\pi j}}\int_{c - j\infty }^{c + j\infty } {G(p)\exp \left( {tp} \right)dp}={\cal L}^{ - 1} \left\{ {G(p);t} \right\}
\end{align}
Let us define the contour given in Fig. \ref{figContour01}, where the value of $c$ is chosen to be at the right of all the singularities in $G(p)$. Note that ${\theta=\tfrac{\gamma^2}{\gamma^2+2\alpha}\in(0,1)}$ by definition.
%\clearpage
Hence, the contour integral along $C$ is given by 
\begin{align}
\label{4eqapp16}
\oint_C {G(p){\mathop{e}\nolimits} ^{\frac{{\beta ^2 p}}{2}} dp}  = \frac{1}{{2\pi j}}\int_{c - j\infty }^{c + j\infty } {G(p){\mathop{e}\nolimits} ^{\frac{{\beta ^2 p}}{2}} dp}  + \oint_{C_\beta  } {G(p){\mathop{e}\nolimits} ^{\frac{{\beta ^2 p}}{2}} dp}.
\end{align}
Using Cauchy-Goursat theorem, we can equivalently express
\begin{align}
\label{4eqapp17}
\oint_C {G(p){\mathop{e}\nolimits} ^{\frac{{\beta ^2 p}}{2}} dp}  = \oint_{\Gamma_0 } {G(p){\mathop{e}\nolimits} ^{\frac{{\beta ^2 p}}{2}} dp}  + \oint_{\Gamma_1 } {G(p){\mathop{e}\nolimits} ^{\frac{{\beta ^2 p}}{2}} dp},
\end{align}
where $\Gamma_0$ and $\Gamma_1$ are closed contours which enclose the singularities at $\{p=0,p=\theta\}$ and $p=1$, respectively. Combining (\ref{4eqapp16}) and (\ref{4eqapp17}), we have
\begin{align}
\label{4eqapp18}
&\underbrace {\frac{1}{{2\pi j}}\int_{c - j\infty }^{c + j\infty } {G(p){\mathop{e}\nolimits} ^{\frac{{\beta ^2 p}}{2}} dp} }_{{\cal L}^{ - 1} \left\{ {G(p);t} \right\}}
 + \oint_{C_\beta  } {G(p){\mathop{e}\nolimits} ^{\frac{{\beta ^2 p}}{2}} dp}  =\\&\nonumber \oint_{\Gamma_0 } {G(p){\mathop{e}\nolimits} ^{\frac{{\beta ^2 p}}{2}} dp}  + \oint_{\Gamma_1 } {G(p){\mathop{e}\nolimits} ^{\frac{{\beta ^2 p}}{2}} dp}.
\end{align}
Since $\left| {G(p)} \right| \le MR^{ - l}$ for some $l>0$ on $C_{\beta}$ as $R \rightarrow \infty$, the integral $\oint_{C_\beta  }$ equals zero. Hence, choosing a contour $\Gamma_0\equiv\Gamma$, we have
\begin{align}
\label{4eqapp19}
I_{m,n,k}^{\beta ,\delta ,\theta} \left( u \right) &=  - \oint_\Gamma  {G(p){\mathop{e}\nolimits} ^{\frac{{\beta ^2 p}}{2}} dp}  \nonumber \\&= \oint_{\Gamma_1 } {G(p){\mathop{e}\nolimits} ^{\frac{{\beta ^2 p}}{2}} dp}  - {\cal L}^{ - 1} \left\{ {G(p)} \right\}_{t = \frac{{\beta ^2 }}{2}}.
\end{align}
Using the residue theorem, we can express
\begin{align}
\label{4eqapp20}
I_{m,n,k}^{\beta ,\delta ,\theta} \left( u \right) = {\rm{Res}}\left\{ {G(p){\mathop{e}\nolimits} ^{\frac{{\beta ^2 p}}{2}} } \right\}_{p = 1}  - {\mathcal{L}}^{{\rm{ - 1}}} \left\{ {{\rm{G(p)}}} \right\}_{{\rm{t = }}\frac{{{\rm{b}}^{\rm{2}} }}{{\rm{2}}}},
\end{align}
where ${\rm{Res}}\left\{ F(p) \right\}_{p = x} $ denotes the residue of $F(p)$ at $p=x$. The calculation of this residue yields
\begin{align}
\label{4eqapp22}
{\rm{Res}}\left\{ {G(p){\mathop{e}\nolimits} ^{\frac{{\beta ^2 p}}{2}} } \right\}_{p = 1}  = \frac{1}{{\left( {1 - \theta } \right)^{n - k} }}e^{\delta \theta u^2 } {\mathop{e}\nolimits} ^{\frac{{\beta ^2 }}{2}}.
\end{align}
To calculate the inverse Laplace transform of $G(p)$, we use partial fraction expansion in (\ref{4eqapp13}) to identify
\begin{align}
\label{4eqapp23}
G(p) &= G_1 (p) + G_2 (p) \nonumber\\&= \frac{1}{{p^{k - (n - m)} }}e^{\frac{{\delta \theta u^2 }}{{p}}} \left( {\frac{{A_0 }}{{p - 1}} + \sum\limits_{l = 1}^{n - k} {\frac{{A_l }}{{\left( {p - \theta } \right)^{l} }}} } \right),
\end{align}
where the constants $A_0$ and $A_l$ are given by 
\begin{align}
A_0&=(1-\theta)^{k-n}\\
A_l&=(1-\theta)^{k+l-n-1}.
\end{align}
Using \cite{Erdelyi1954}, we find an expression for the inverse Laplace transforms as
\begin{align}
\label{4eqapp24}
{\cal L}^{ - 1}& \left\{ {G_1 (p)} \right\}_{t = \frac{{\beta ^2 }}{2}}  = \\&A_0 \left( {\frac{{\beta ^2 	}}{2}} \right)^{k + m - n} \tilde{\Phi} _3 \left( {1,k + m - n + 1;\tfrac{{\beta ^2 }}{2},\tfrac{{\delta \theta u^2\beta ^2 }}{2}} \right)\nonumber,
\end{align}
and
\begin{align}
\label{4eqapp25}
&{\cal L}^{ - 1} \left\{ {G_2 (p)} \right\}_{t = \frac{{\beta ^2 }}{2}}  = \\&\nonumber\sum\limits_{l = 1}^{n - k} {A_l} \left( {\frac{{\beta ^2 }}{2}} \right)^{k + m - n + l - 1} \tilde{\Phi} _3 \left( {l,k + m - n + l; \tfrac{{\theta\beta ^2 }}{2},\tfrac{{\delta \theta u^2\beta ^2 }}{2}} \right),
\end{align}
where $\tilde{\Phi_3}(b,c,w,z)$ is the regularized confluent hypergeometric function of two variables, defined as a normalized version of the confluent hypergeometric function of two variables given in \cite[9.261.3]{Gradstein2007}, i.e.
\begin{align}
\label{4eqapp26}
\tilde{\Phi_3}(b,c,w,z)\triangleq \frac{\Phi_3(b,c,w,z)}{\Gamma(c)}.
\end{align}
Finally, combining (\ref{4eqapp22}), (\ref{4eqapp24}) and (\ref{4eqapp25}) yields the desired expression for (\ref{4eqapp01}).

\section{Generalized $Q$-functions in communication theory}
\label{AppQ}

In this appendix, we introduce some fundamentals on the generalized $Q$-functions used in this paper, which have played a key role in communication theory for almost 70 years. The Marcum $Q$-function is named after Jess Ira Marcum (born Marcovitch \cite{blackjack2005}), who first introduced this nomenclature in 1948 when developing a statistical theory of target detection by pulsed radar \cite[eq. 16]{Marcum1948} as
\begin{align}
\label{eqQ01}
Q({\alpha,\beta})=\int_{\beta}^{\infty}{x \exp{\left(-\frac{x^2+\alpha^2}{2}\right)}I_0(\alpha x)dx},
\end{align}
where $I_0(\cdot)$ is the modified Bessel function of the first kind and order zero. A more general form of this integral was also included in Marcum's original report \cite[eq. 49]{Marcum1948}, which is usually referred to as generalized Marcum $Q$-function. This function can be expressed as
\begin{align}
\label{eqQ02}
Q_M({\alpha,\beta})=\frac{1}{\alpha^{M-1}}\int_{\beta}^{\infty}{x^M\exp{\left(-\frac{x^2+\alpha^2}{2}\right)}I_{M-1}(\alpha x)dx},
\end{align}
where $I_{M-1}(\cdot)$ is the modified Bessel function of the first kind and order $M-1$. We see that (\ref{eqQ02}) reduces to (\ref{eqQ01}) for $M=1$, and hence it is usual to include the subindex in $Q_1(\alpha,\beta)$ when using the function in (\ref{eqQ01}). Interestingly, the generalized Marcum $Q$-function can be expressed in terms of the standard Marcum $Q_1$ function, and a finite number of modified Bessel functions as \cite[eq. 4.81]{Simon2005}
\begin{align}
\label{eqQ02b}
Q_M({\alpha,\beta})= Q_1(\alpha,\beta)+ \exp{\left(-\frac{\alpha^2+\beta^2}{2}\right)} \sum_{n=1}^{M-1}\left(\frac{\beta}{\alpha}\right)^n I_{n}(\alpha \beta).
\end{align}

This family of functions is relevant not only in communication theory but in general statistics, since the generalized Marcum $Q$-function represents the ccdf of the non-central chi-squared distribution (i.e., the squared norm of a Gaussian vector) \cite{Simon2006}.

Later in the 70's, Albert H. Nuttall introduced a more general form of this function \cite[eq. 86]{Nuttall1974} as
\begin{align}
\label{eqQ03}
Q_{M,N}({\alpha,\beta})=\int_{\beta}^{\infty}{x^M\exp{\left(-\frac{x^2+\alpha^2}{2}\right)}I_{N}(\alpha x)dx},
\end{align}
which is usually referred to as Nuttall $Q$-function \cite{Simon2002}. Letting $M=N+1$, the Nuttall $Q$-function relates to the generalized Marcum $Q$-function as
\begin{align}
\label{eqQ04}
Q_{N+1,N}({\alpha,\beta})=\alpha^N Q_{N+1}(\alpha,\beta),
\end{align}
whereas the standard Marcum $Q$-function is given by $Q_{1,0}({\alpha,\beta})=Q_{1}(\alpha,\beta)$. As previously stated in (\ref{5eqapp02}), the Nuttall $Q$-function can be expressed in terms of a finite number of modified Bessel functions and generalized Marcum $Q$-functions if the sum of the indices $M+N$ is an odd number.
  
\bibliographystyle{ieeetr}
\bibliography{BivEigenvalues}

\begin{IEEEbiographynophoto}{F. Javier Lopez-Martinez} (S'05, M'10) received the M.Sc. and Ph.D. degrees in Telecommunication Engineering in 2005 and 2010, respectively, from the University of Malaga (Spain). He joined the Communication Engineering Department at the University of Malaga in 2005, as an associate researcher. In 2010 he stays for 3 months as a visitor researcher at University College London. He is the recipient of a Marie Curie fellowship from the UE under the “U-mobility” program at University of Malaga. Within this project, between August 2012-2014 he held a postdoc position in the Wireless Systems Lab (WSL) at Stanford University, under the supervision of Prof. Andrea J. Goldsmith. He's now a Postdoctoral researcher at the Communication Engineering Department, Universidad de Malaga.
\\
\indent He has received several research awards, including the best paper award in the Communication Theory symposium at IEEE Globecom 2013, and the IEEE Communications Letters Exemplary Reviewer certificate in 2014. His research interests span a diverse set of topics in the wide areas of Communication Theory and Wireless Communications: stochastic processes, random matrix theory, statistical characterization of fading channels, physical layer security, massive MIMO and mmWave for 5G.
\end{IEEEbiographynophoto}

\begin{IEEEbiographynophoto}{Eduardo Martos-Naya} received the M.Sc. and Ph.D. degrees in  telecommunication engineering from the University of Malaga, Malaga, Spain, in 1996 and 2005, respectively. In 1997, he joined the Department of Communication Engineering, University of Malaga, where he is currently an Associate Professor. His research activity includes digital signal processing for communications, synchronization and channel estimation, and performance analysis of wireless systems. Currently, he is the leader of a project supported by the Andalusian regional  Government on cooperative and adaptive wireless communications systems.
\end{IEEEbiographynophoto}

\begin{IEEEbiographynophoto}{Jos\'e F. Paris}
received his MSc and PhD degrees in Telecommunication Engineering from the University of M\'alaga, Spain, in 1996 and 2004,
respectively. From 1994 to 1996 he worked at Alcatel, mainly in the development of wireless telephones. In 1997, he joined the
University of M\'alaga where he is now an associate professor in the Communication Engineering Department. His teaching activities
include several courses on signal processing, digital communications  and acoustic engineering. His research interests are related to wireless
communications, especially channel modeling and performance analysis. Currently, he is the leader of several projects supported by
the Spanish Government and FEDER on wireless acoustic and electromagnetic communications over underwater channels. In 2005,
he spent five months as a visitor associate professor at Stanford University with \mbox{Prof. Andrea J. Goldsmith.}
\end{IEEEbiographynophoto}

\begin{IEEEbiographynophoto}{Andrea Goldsmith} (S'90-M'93-SM'99-F'05) is the Stephen Harris professor in the School of Engineering and a professor of Electrical Engineering at Stanford University. She was previously on the faculty of Electrical Engineering at Caltech. Dr. Goldsmith co-founded and served as CTO for two wireless companies: Accelera, Inc., which develops software-defined wireless network technology for cloud-based management of WiFi  access points, and  Quantenna Communications, Inc., which develops high-performance WiFi chipsets. She has previously held industry positions at Maxim Technologies, Memorylink Corporation, and AT\&T Bell Laboratories. She is a Fellow of the IEEE and of Stanford, and has received several awards for her work, including the IEEE ComSoc Armstrong Technical Achievement Award,  the National Academy of Engineering Gilbreth Lecture Award, the IEEE ComSoc and Information Theory Society joint paper award, the IEEE ComSoc Best Tutorial Paper Award, the Alfred P. Sloan Fellowship, and the Silicon Valley/San Jose Business Journal’s Women of Influence Award. She is author of the book ``Wireless Communications'' and co-author of the books ``MIMO Wireless Communications'' and “Principles of Cognitive Radio,” all published by Cambridge University Press, as well as inventor on 25 patents. She received the B.S., M.S. and Ph.D. degrees in Electrical Engineering from U.C. Berkeley.
\\
\indent Dr. Goldsmith has served as editor for the IEEE Transactions on Information Theory, the Journal on Foundations and Trends in Communications and Information Theory and in Networks, the IEEE Transactions on Communications, and the IEEE Wireless Communications Magazine as well as on the Steering Committee for the IEEE Transactions on Wireless Communications. She participates actively in committees and conference organization for the IEEE Information Theory and Communications Societies and has served on the Board of Governors for both societies. She has also been a Distinguished Lecturer for both societies, served as President of the IEEE Information Theory Society in 2009, founded and chaired the Student Committee of the IEEE Information Theory Society, and chaired the Emerging Technology Committee of the IEEE Communications Society. At Stanford she received the inaugural University Postdoc Mentoring Award, served as Chair of Stanford’s Faculty Senate in 2009, and currently serves on its Faculty Senate, Budget Group, and Task Force on Women and Leadership. 
Biography text here.
\end{IEEEbiographynophoto}

\end{document}